\pdfoutput=1
\documentclass[11pt]{article}

\usepackage[affil-it]{authblk} 
\usepackage[authoryear]{natbib}
\usepackage{gensymb}
\usepackage{amsmath}
\usepackage{amssymb}
\usepackage{graphicx}
\usepackage{float}
\usepackage{booktabs}
\usepackage[a4paper, top=1in,bottom=1in,left=1in,right=1in,marginparwidth=0cm]{geometry}
\usepackage[title]{appendix}
\usepackage{pdflscape}
\usepackage{longtable}
\usepackage{setspace}
\usepackage{adjustbox}
\usepackage{array}
\usepackage{caption}
\usepackage{chngcntr}
\usepackage{mathtools}
\usepackage{lineno}
\usepackage{makecell}
\usepackage{hyperref}
\hypersetup{colorlinks=true,allcolors=blue}
\usepackage{longtable}
\usepackage{array}

\newcolumntype{R}[2]{%
    >{\adjustbox{angle=#1,lap=\width-(#2)}\bgroup}%
    l%
    <{\egroup}%
}
\newcommand*\rot{\multicolumn{1}{R{45}{1em}}}

\newcommand{\citepos}[1]{\citeauthor{#1}'s \citeyear{#1}}
\renewcommand{\eqref}[1]{Eq.\,\ref{#1}}

\title{An assessment of Alberta's strategy for controlling mountain pine beetle outbreaks}

\author[1,*]{Evan C. Johnson}

\author[2,3]{Mark A. Lewis}
\affil[1]{Mathematical and Statistical Sciences; University of Alberta; Edmonton, Alberta, Canada}
\affil[2]{Department of Mathematics and Statistics; University of Victoria; Victoria, British Columbia, Canada}
\affil[3]{Department of Biology; University of Victoria; Victoria, British Columbia, Canada}
\affil[*]{Corresponding author: Evan Johnson, ecjohns1@ualberta.ca}

\date{} 

\begin{document}

\maketitle 

\newpage

\tableofcontents 

\newpage 

\section*{Abstract} 

The Canadian province of Alberta spent over 500 million dollars on controlling mountain pine beetle populations, but did it work? Using a statistical modeling framework coupled with long-term field data, we examined how direct control measures, severe winters, and host-tree depletion shaped the trajectory of Alberta’s mountain pine beetle outbreak between 2009 and 2020. Simulations suggest that control efforts reduced total tree mortality by 79\% (95\% predictive interval: 58--89\%) and prevented 1.8 (0.91--4.1) trees per hectare from being killed from 2010--2020. Although cold winters had little effect on overall damage, they acted synergistically with control to end the outbreak, causing population collapse circa 2020. This synergy supports a “wait it out” strategy of mountain pine beetle management, where moderate control effort is applied until an extreme weather event delivers the final blow. Any effects of host-tree depletion via beetle attack were negligible. From an economic perspective, removing one infestation tree --- at an approximate cost of 320 CAD --- prevented the loss of roughly six (2.6--15) trees, demonstrating the potential for long-term cost-effectiveness. Our results further indicate that future outbreaks may vary widely in severity due to environmental stochasticity, with potential damage in a no-control scenario ranging from 0.41 to 9.7 trees per hectare killed (over a hypothetical 11-year period). An alternative model predicts an even wider range of outcomes: 1--40 trees per hectare. These findings highlight not only the potential of sustained control efforts in mitigating forest pest outbreaks, but also the inherent uncertainty in long-term ecological forecasting.

\newpage

\section{Introduction} \label{Introduction}

The mountain pine beetle (MPB; \textit{Dendroctonus ponderosae} Hopkins) outbreak in British Columbia (BC) during the early 2000s was unprecedented in its scale and severity \citep{taylor2006forest}. Beginning in the late 1990s, a combination of factors --- specifically favorable climatic conditions and abundant mature pine forests --- led to an explosion of beetle populations across BC's interior forests \citep{taylor2003disturbance, carroll2006impacts, alfaro2009historical, creeden2014climate}. Management efforts in BC were minimal. The \textit{BC Ministry of Forests} increased the Annual Allowable Cut to salvage dead trees, and in rare cases, to suppress incipient outbreaks in particular timber supply areas \citep{forest2007timber}. However, given the perfect storm of conditions (i.e., drought, warmer winters, fire suppression leading to over-mature pine stands) it's questionable whether more intensive management would have been effective. By 2005, the infestation had affected over 8.7 million hectares of pine forests, resulting in the loss of approximately 400 million $\text{m}^3$ of timber \citep{BC2006mountain}. This ``hyperepidemic'' facilitated a northeastward range expansion, following the prevailing winds --- since only a small percentage of beetles engage in long-distance dispersal, large populations in the outbreak epicenter ensured that enough beetles landed in new areas to overwhelm tree defenses \citep{johnson2006allee, bleiker2019risk}. MPB spread into Northern BC, the Yukon, and the Northwest Territories; most notably, they crossed the Rocky Mountains --- a former geographical barrier --- into Alberta \citep{jackson2008radar}.

\begin{figure}[H]
\centering
\includegraphics[scale = 1]{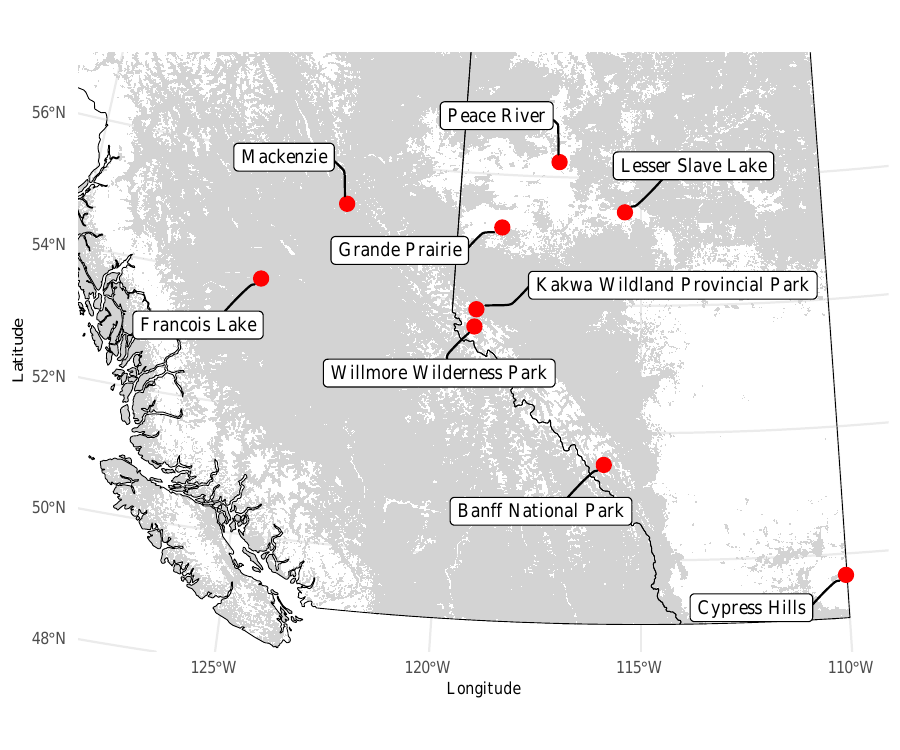}
\caption{Key locations in our telling of the mountain pine beetle range expansion. The grey background shows areas with pines (live aboveground biomass $> 1 \, \text{Mg} \, \text{ha}^{-1}$).}
\label{fig:Intro_locations}
\end{figure}

The history of Mountain Pine Beetle infestations in Alberta can be divided into two distinct periods: sporadic occurrences before the 2000s and a dramatic range expansion driven by multiple long-distance dispersal events from British Columbia. Historical information in this introduction draws from \citep{brett2024history} with key locations shown in Figure \ref{fig:Intro_locations}. MPB had been recorded in Alberta's western parks and Cypress Hills (on the borders of Alberta and Saskatchewan) in previous decades. A 1981 survey, conducted by the Canadian Forest Service to reveal the extent of MPB dispersal, found MPB in dozens of locations across Alberta's prairies; however, MPB was invariably found in isolated pine refugia like arboretums and shelterbelts. The first signs of the new expansion were observed in the Willmore Wilderness Park in 1999, with small pockets of MPB-killed trees detected along the Upper Smoky River. By 2005, infestations had appeared in several rocky mountain valleys and had reached as far north as the Kakwa Wildland Provincial Park; some infestations appeared in the Rocky Mountain foothills, south of Grand Prairie. In 2006, a large beetle migration led to a large range expansion, reaching north of the Peace river near the BC border (57°45' latitude) and eastward to Lesser Slave Lake (115° 30’ longitude). Genetic studies show these expansions in 2005 and 2006 occurred through separate long-distance dispersal events, originating from central BC populations near Mackenzie and Francois Lake, respectively \citep{samarasekera2012spatial}.

Alberta's response to the 2000s outbreak was swift and multifaceted, focusing on both control and monitoring efforts. The Alberta Sustainable Resource Development took the lead in implementing control measures, which included a combination of detection surveys, pheromone baiting, and removing infested trees. In the Willmore Wilderness Area alone, over 10,000 infested trees were treated by May 2006. The province also collaborated with Parks Canada to implement a two-zone management approach in Banff National Park, establishing a monitoring zone where prescribed fires were used to reduce MPB habitat, and a management zone where more intensive control measures were applied.

Monitoring efforts intensified after the 2006 range expansion, with both aerial and ground surveys conducted annually. These surveys revealed that over 2.5 million trees had been attacked, a discovery that prompted Alberta to shift its management strategy from eradication to suppression and prevention. The province established a \textit{Leading Edge Zone} to encompass top-priority areas, including areas deemed most likely to drive eastward expansion.

Alberta's mountain pine beetle management strategy consisted of three control methods: Level 1 treatment, Level 2 treatment, and the \textit{healthy pine strategy} \citep{government2007alberta}. Level 1 treatment, the most widely applied method, involved the removal of individual infested trees. Level 2 control consisted of larger-scale harvesting operations (usually clear-cut logging) targeting heavily infested stands.  The healthy pine strategy, implemented in 2006, allowed forestry companies to increase their logging quotas if they agreed to preemptively remove stands that were either currently susceptible (i.e., climactically suitable for MPB with older pines) or would soon become susceptible to MPB. The strategy aimed to reduce susceptible pine by over 75\% over a 20-year period.

This paper focuses on Level 1 treatment, both because it is the most costly method in Alberta's MPB management portfolio, and because the other control methods were either less effective or could not be properly evaluated. The limited effect of Level 2 treatment (harvest) was demonstrated with \textit{MPB Online}, a stochastic simulation model developed by forestry consulting company FORCORP.  Initial scenario modeling showed that temporary Level 1 control reduced affected pine volume by 31\% over a 20-year period, while temporary Level 2 control  only achieved a 5\% reduction \citep{forcorp2014mpb}. An updated version of the model found an even larger difference \citep{forcorp2020mpb}. Level 2 control was likely limited by logistics: many infested stands were inaccessible due to limited logging roads, and forestry companies typically plan their operations 3--5 years in advance, making it difficult to rapidly respond to new MPB outbreaks (\textit{Mike Undershultz}, personal communication).

The effectiveness of the healthy pine strategy is unknown. Such an assessment would require up-to-date harvest data from various forestry companies, which is not currently available  (\textit{Mike Undershultz}, personal communication). Given that level 2 treatment is of limited effect and the healthy pine strategy remains unevaluated, throughout this paper we will often use the word ``control'' as a shorthand for level 1 treatment specifically.

Level 1 treatment involved a combination of surveys and sanitation \citep{government2016mountain}. The process began with helicopter surveys to locate clusters of newly killed (red-topped) trees, which indicated potential MPB infestations. Once these areas were identified, ground crews conducted intensive surveys to locate currently infested (green-attack) trees. The ground surveys used a concentric circle method, where crews searched within a 50-meter radius of identified red-attack trees. The survey plot could be extended under certain conditions: if a green-attack tree was found beyond the inner circle with a 25 m radius, this triggered a mini-survey on the margins of the 50 m plot; if additional green-attack trees were found there, an adjacent but non-overlapping 50 m plot was established. Nearly all detected green-attack trees were flagged and removed. The primary removal/sanitation method was known as ``cut-and-burn'', where infested trees were felled and then burned on-site.

The effectiveness of level 1 treatment at the landscape scale was compromised by long-distance beetle dispersal, resource constraints, and environmental variability. Field crews are nearly perfect at detecting green-attack trees within concentric surveys (98.5\%; accuracy; Caroline Whitehouse pers. comm. in \citealp{bleiker2023suitability}), but many beetles were outside of survey areas. Approximately half of the beetles establish themselves more than 50 meters away from their natal trees \citep{johnson2024stratified}, beyond the typical survey radius. Resource constraints further complicate the situation, as field crews were unable to visit all locations with red-topped trees. Instead, sites were prioritized based on MPB population forecasts, the number of red-topped trees, a stand susceptibility index, and forest connectivity \citep{government2007alberta, hodge2017strategic}. These limitations resulted in an overall \textit{control efficacy} --- defined as the proportion of infestations successfully removed with level 1 treatments --- of approximately 50\%. \citet{carroll2017assessing} observed significant year-to-year variability in control efficacy (range: 38--68\%), putatively due to ephemeral weather conditions affecting MPB population dynamics within Alberta, or idiosyncratic long-distance immigration from British Columbia.

The long-term effectiveness of MPB control may depend on the interaction between direct control and extreme cold events. Winter temperatures below -35°C cause significant mortality, and temperatures below -40°C cause near 100\% mortality \citep{wygant1940effects}. The beetles build up their cold hardiness and lose it gradually in spring, leaving them vulnerable to cold snaps in the range of -20°C-- -30°C, in autumn or spring. \citep{safranyik1991unseasonably, regniere2007modeling}. This sensitivity offers suggests a ``wait it out'' strategy. By applying Level 1 treatments consistently, forest managers could limit outbreak growth, effectively buying time until a sufficiently cold-weather event occurs. Such an event could reduce beetle populations below their Allee threshold --- the minimum population size required for positive growth \citep{taylor2012allee}. MPB exhibits a strong Allee effect, where per capita growth rates become negative at low population densities \citep{raffa1983role, waring1983physiological, boone2011efficacy}. This characteristic makes short-term population reduction a viable strategy for long-term outbreak control and could justify continued funding for MPB management in the face of imperfect control control measures.

Several previous studies have explored the efficacy of control. A cellular-automaton model called \textit{MPBSpread} \citep{carroll2017assessing} implied that Alberta's slow-the-spread strategy, which includes level 1 and level 2 treatments, reduced the area colonized by MPB to approximately 70\% of that predicted under a do nothing scenario. Model outcomes were particularly sensitive to early detection and the amount of level 1 control applied --- doubling the area treated with level 1 control significantly reduced colonized area compared to the standard slow-the-spread approach. These findings are supported by an earlier model \citep{carroll2006direct}, which implied that control efficacy must exceed 67\% to suppress MPB populations increasing at a modest rate of threefold annually. Given that real-world control efficacy hovers around 50\%, level 1 treatments as applied are unlikely to fully suppress MPB outbreaks, though they may slow their progression. Finally, there is the previously mentioned \textit{MPB online} model developed by FORCORP. While we mention this model for completeness, its long-term predictions are dubious given that the model does not represent MPB dispersal over 1 km, and more broadly, the model's quality cannot be evaluated because a complete model description is not publically avaiable (to our knowledge).

While previous models provide valuable initial results, their quantitative predictions are can be improved upon with respect to elements that ensure robustness. These are 1) transparent methods for model calibration, 2) in-depth analysis of factors affecting model uncertainty and 3) use of models of intermediate complexity so as to have adequate realism but avoid overcomplexity. The 2006 model \citep{carroll2006direct} used a single parameter (MPB reproduction rate) to demonstrate the importance of early, sustained control efforts. Such strategic models (\textit{sensu} \citealp{holling1966strategy}) are not suited for making quantitative statements about the Alberta outbreak. The 2016 \textit{MPBSpread model}, though more realistic, contains many parameter values and constitutive submodels given without full justification. A follow-up paper (which used the same model) indicated that at least some of the model's 20 parameters were manually adjusted to align with survey data from Alberta, raising concerns about overfitting \citep[p.15]{carroll2020alternative}. For these reasons, it is critical to undertake an independent assessment that explicitly addresses model calibration, uncertainty analysis, and model complexity.

\begin{enumerate}

\item \textbf{What ended the outbreak?} We examine the relative importance of control efforts, cold winters, and host depletion in terminating the MPB outbreak. While all three mechanisms have been proposed (Janice Cooke qtd. in \citealp{cbc_mountain_pine_beetle}), their respective contributions remain unclear. Cold winters are known to cause population crashes, while control efforts at this scale are unprecedented. Host depletion, though less likely given Alberta's low infestation densities, remains a possibility. We also explore how control efforts might interact with cold events, potentially ``buying time'' until severe overwintering mortality occurs.

\item \textbf{How many trees did control efforts save?} We use simulations to compare control and no-control scenarios, assessing both the absolute number of trees saved and the percentage of total damage mitigated. These metrics provide measures of absolute and relative control effectiveness in the long term. Our analysis covers 2009 to 2020, a period bounded by two significant events: a major MPB immigration from British Columbia in 2009, which effectively reset Alberta's landscape of infestations, and the outbreak's approximate end in 2020.

\item \textbf{What range of control effectiveness should we expect in future outbreaks?} To address this question, we model potential future outbreaks by allowing winter temperatures and other environmental factors (represented as year effects) to vary from year to year. Given the large interannual fluctuations in MPB population densities, we anticipate a wide range of outcomes with respect to long-term control effectiveness. At one extreme, some future outbreaks might result in substantial tree mortality despite control efforts. On the other hand, moderate control measures could entirely suppress outbreaks.

\item \textbf{How does the cost-effectiveness of MPB control vary with control intensity?} Theory suggests that increasing control efforts could yield disproportionate benefits by pushing beetle populations below the Allee threshold, swiftly ending outbreaks and greatly reducing future expenses. We calculate the cost per hectare of level 1 treatment at different levels of control efficacy. Additionally, we calculate the efficiency --- measured here as trees saved per 100 CAD --- under Alberta's actual control regime; this enables stakeholders to compare control costs against the economic and ecological benefits of retaining trees.

\end{enumerate}

These questions are answered sequentially in labeled subsections of the \textit{Results}. We aim to contribute to more informed decision-making in forest management, to help balance ecological and economic concerns in the face of highly uncertain population dynamics.

Importantly, this study considers multiple sources of uncertainty. Parameter uncertainty is handled by simulating with a range of reasonable parameter values, providing a more complete view of possible outbreak trajectories. Environmental stochasticity enables prospective simulations of future outbreaks under varying environmental conditions. An analysis of alternative model structures (Appendix \ref{Alternative models}) demonstrates that quantitative predictions are heavily dependent on model structure --- a crucial consideration given our imperfect understanding of landscape-scale MPB dynamics.

\section{Methods} \label{Methods}

\subsection{Overview} \label{Overview}

Using an empirically calibrated model of mountain pine beetle dynamics, we simulate the Alberta outbreak under different scenarios, e.g., with and without level 1 treatment. While conceptually simple, this methodology has many moving parts, including multiple data sources (Section \ref{Data}), multiple submodels (Section \ref{Model description}), model justification (Section \ref{Model justification}), and the simulation procedure (Section \ref{Simulations}). Counterfactual simulations can be used to reveal the factors responsible for the salient features of the Alberta outbreak, specifically the collapse of the outbreak around 2020, and the total number of killed trees over an 11-year period (2010--2020). For example, if population densities never collapse when MPB control is turned off, we can say that control was necessary for ending the outbreak. If a collapse only occurs when control \textit{and} severe winters are turned off, then those two factors are jointly necessary.

The model predicts beetle infestations through three main life-history processes: overwintering survival (based on weather data), dispersal (modeled with a Student's $t$-distribution), and local population dynamics (modeled with a zero-inflated negative binomial distribution, or ZINB). Space is discretized into 5x5 km cells within a 2D lattice, with each cell serving as a single observation. The ZINB structure reflects mountain pine beetle biology: the zero-inflation captures both clustered dispersal and MPB's well-known \textit{Allee effect}, where small populations fail to overcome tree defenses, while the negative binomial component accounts for variable infestation counts due to clustered dispersal and any number of un-modeled factors (e.g., within-cell habitat heterogeneity). The model uses three predictors: beetle pressure (i.e., beetles arriving after their dispersal phase), pine volume, and cumulative tree mortality. The model also includes year effects to account for unexplained annual variation in beetle productivity, and Gaussian processes to address residual spatial autocorrelation. 

Our focus on model-justification, validation, and comparisons contrasts with previous MPB models, which often make biologically inaccurate assumptions (for further discussion of these issues, see Johnson et al., \citeyear{johnson2024explaining} \& \citeyear{johnson2024stratified}). To explore how different modeling choices may affect our conclusions, we analyze two additional models in Appendix \ref{Alternative models}. We focus on a single model in the main text, both for simplicity of presentation, and because this model demonstrated superior predictive ability (Appendix \ref{Model comparisons}). 


Understanding the MPB life cycle is essential for understanding data collection. Adult MPB emerge in July or August of each year, leaving their dead natal trees to locate new hosts. This dispersal / host-finding period lasts 2--6 weeks \citep{bleiker2016flight}. Beetles attack \textit{en masse}, using aggregation pheromones and host volatiles to overwhelm the resin defenses of pine trees \citep{raffa2001mixed}. Adults build egg galleries under the bark and subsequently die, ending their univoltine life cycle. Successfully attacked trees die within months, and their needles turn red about a year later. A more detailed description of MPB life history is given by \citet{safranyik2006biology}.

\subsection{Data} \label{Data}

Mountain pine beetle infestation data comes from \textit{Heli-GPS surveys} and \textit{concentric ground surveys}. Each year in September or October, Alberta's Department of Forestry and Agriculture conducts helicopter surveys to locate clusters of \textit{red-topped trees}: trees first infested the previous year. The locations of these clusters are recorded (hence the name \textit{Heli-GPS surveys}), and the number of infested trees within a cluster is counted. Since MPB typically disperse short distances \citep{safranyik1992dispersal}, field crews are dispatched to these GPS points to search for \textit{green-attack trees}: newly infested trees without red needles. The vast majority of discovered green-attack trees are then ``sanitized'' by felling and either burning or chipping.

Alberta's MPB surveys produce high-quality data with minimal measurement errors. The Heli-GPS surveys accurately count the number of infestations (within $\pm 10$ trees) for 92\% of infestation clusters \citep{nelson2006large}. Clusters of red-topped trees are located with a positional accuracy of $\pm 30$ meters \citep{government2016mountain}. The concentric ground surveys are even more accurate, with a 98.5\% detection rate within the surveyed area \citep{bleiker2019risk}.

Our analysis incorporates several additional data sources. We use estimates of pine volume calculated from the extended Alberta Vegetation Inventory (AVI; \citealp{alberta_vegetation_inventory_2022}), which is a standardized digital forest inventory product derived from high-resolution aerial imagery, silvicultural records, and field data collected by forestry companies. Photo interpreters delineate forested areas and estimate species composition (crown closure percentages) within each polygon. The conifer volume attribute (units: $\text{m}^3 \text{ha}^{-1}$) is estimated with a model that incorporates stand height, age, and species-specific site index equations. We derive the total pine volume within a cell (units: $\text{m}^3$) by multiplying the conifer volume attribute, times the proportion of conifers that are pines, times the total forested hectares. 

The model also uses the probability of overwintering survival,  which is calculated by taking weather station temperature data, spatially imputing it with the \textit{BioSIM} program \citep{regniere2014biosim}, then feeding it into the predictive model of \citet{regniere2007modeling}. While this model is sophisticated --- daily temperatures influence the slow accumulation of beetle cold tolerance, thus modeling the effects of both unseasonable cold snaps and extreme cold --— results indicate that overwintering survival is roughly proportional to minimum winter temperature (Fig. \ref{fig:Psurv_vs_Tmin}). Finally, we used pine population-genetic data from \citet{cullingham2012characterizing} to exclude cells that predominantly contained jack pine or jack-lodgepole hybrids, as there is recent evidence that Jack pine forests are less susceptible to MPB. \citep{srivastava2023dynamic, johnson2024mountain}.

\subsection*{Data preparation} \label{data_prep}

The raw Heli-GPS and ground survey data contain GPS coordinates and the number of trees. We rasterized this point data into 5$\times$5 km cells; while admittedly coarse, this resolution allowed us to include the computationally intensive Gaussian processes. 

The total number of infested trees in cell $x$ and year $t$, denoted $I_t(x)$, is calculated as 
\begin{equation}
I_t(x) = m_{t}(x) + r_{t+1}(x),
\end{equation}
where $m_{t}(x)$ is the number of sanitized green-attack trees in the focal year, and $r_{t+1}(x)$ is the number of red-topped trees observed in the following year. While $I_t(x)$ is a suitable response variable (i.e., that which is predicted), it is not a suitable predictor (i.e., a model input), because it contains sanitized trees that cannot contribute to future infestations. Therefore, we define a slightly modified input variable: $I_{t}^{*}(x) = I_t(x) - m_t(x)$.

\begin{figure}[H]
\centering
\includegraphics[scale = 0.8]{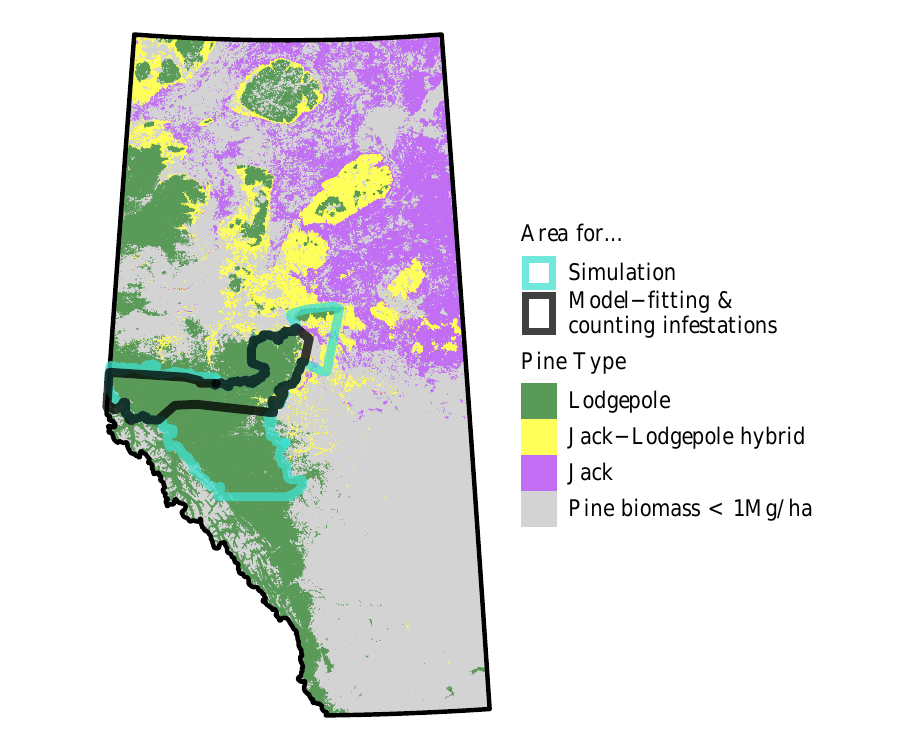}
\caption{Areas in Alberta that are used for model-fitting and simulations. The pine species data comes from \citet{cullingham2012characterizing}. The pine biomass data comes from \citet{beaudoin2014mapping}. Here we use biomass instead of AVI-based pine volume estimates, because the latter is not available for the entirety of Alberta.}
\label{fig:study_areas}
\end{figure}

Our analysis required data subsetting based on both biological knowledge (e.g., habitat suitability, immigration events) and practical limitations (e.g., survey coverage). Both model-fitting and simulations only consider the years 2009--2020. A major MPB immigration from BC in 2009 \citep{carroll2017assessing} reset the spatial distribution of infestations in western Alberta, and 2020 marked both the approximate end of the outbreak. We do not attempt to predict the 2009 infestations, since we lack both the data and atmospheric models to characterize extreme long-distance dispersal from central BC. We attempt to subset to cells where MPB infestations are feasible: those with pine volume exceeding 28,250 $\text{m}^3$ (only 2\% of infestations occur below this threshold) and with pure Lodgepole pine populations (excluding cells where Jack pine ancestry is less than 10\% \textit{sensu }\citealp{cullingham2012characterizing}). For model-fitting, we further subset to the intersection of surveyed areas across years (Fig \ref{fig:study_areas}). Our simulations covered a broader spatial extent; however, when calculating summary statistics from simulated data, we subset to the smaller spatial extent to ensure fair comparisons with real data.

\subsection{Model description} \label{Model description}

The probability of potentially observing MPB infestations in cell $x$ during year $t$ is denoted $\pi_t(x)$. Note that this presence/absence submodel only generates the \textit{potential} for a non-zero number of infestations, as the count model may still generate zeros. The probability $\pi_t(x)$ is a linear combination of predictors on the logit scale,
\begin{equation} \label{eq:pi}
\pi_t(x) = \text{logit}^{-1}\left(\gamma_0 + \gamma_B B_t(x) + \gamma_V V(x) + \gamma_K K_t(x) + \tau_{\gamma} \eta_t + \sigma_{\gamma} f_t(x) \right),
\end{equation}
where:
\begin{itemize}
\item $B_t(x)$ is the natural log of beetle pressure at location $x$ in year $t$. The beetle pressure is calculated by multiplying last year's infestations by the probabilities of overwintering survival (element-wise across all cells), and then applying the dispersal submodel (described below). Thus beetle pressure is proportional to the expected number of attacking beetles. 

\item $V(x)$ is the natural log of pine volume at location $x$. Note that there is no time-dependence here (justified in Section \ref{Model justification}). 

\item $K_t(x)$ represents the natural log of the cumulative number of killed trees plus 1 (to avoid $-\infty$), at location $x$, from 2006 to year $t$. More explicitly, $K_t(x) = \log\left(1 + \sum_{s=0}^{t-1} I_s(x)\right)$.

\item $\eta_t$ are year effects in the presence/absence model

\item $f_t(x)$ is a realization of a Gaussian process.

\item $\gamma_0, \gamma_B, \gamma_V,$ and $\gamma_K$ are model coefficients.
\end{itemize}

The year effects and Gaussian processes have hierarchical structures. Specifically, the year effect is distributed via the standard normal (i.e., $\eta_t \sim \text{Normal}\left(0, 1 \right)$), but is scaled by the hyperparameter $\tau_{\gamma}$. The realization of the Gaussian process for all cells, denoted $\overrightarrow{f_t(x)}$, is drawn from a multivariate normal, 
\begin{equation}
     \overrightarrow{f_t(x)} \sim \text{Normal}\left(\overrightarrow{0}, \Sigma \right),
\end{equation}
where the covariance matrix is given by the squared exponential kernel with a single hyperparameter: the length-scale $L$. The covariance between cells $x$ and $y$, whose midpoints are separated by $\text{dist}(x,y)$ km in Euclidean space, is given by 

\begin{equation} \label{eq:cov}
    \Sigma(x,y) = \exp\left( - \, \frac{\text{dist}(x,y)^2}{2 L^2} \right).
\end{equation}

The count submodel has a similar structure to the presence/absence submodel. The mean of the negative binomial distribution is 

\begin{equation} \label{eq:mu}
\mu_t(x) = \exp\left[ \beta_0 + \beta_B B_t(x) + \beta_V V(x) + \beta_K K_t(x) + \tau_{\beta} \eta_t + \sigma_{\beta} f_t(x) \right], 
\end{equation}

where: 
\begin{itemize}
\item The predictors are the same ($B, V, K$) but the coefficients are different ($\beta_0, \beta_B, \beta_V,$ and $\beta_K$).

\item The $\eta_t$ are the same year effects that are used in the presence/absence submodel, just scaled differently (by $\tau_{\beta}$). 

\item $f_t(x)$ is the same realization of the Gaussian process that is used in the presence/absence submodel, just scaled differently (by $\sigma_{\beta}$). 

\end{itemize}

The conditional number of infestations follows a negative binomial distribution, parameterized by a mean and the dispersion/size parameter $k$. The probability mass is 
\begin{equation} \label{eq:count}
f(I_t(x)) = \text{NB}\left(I_t(x) \; | \; \mu_t(x) ,  k \right).
\end{equation}
To be more precise, $\text{NB}$ denotes the probability mass function, 
\begin{equation} \label{eq:NB}
\text{NB}(y \; | \; \mu, k) = \binom{y + k - 1}{y} \left(\frac{\mu}{\mu + k}\right)^y \left(\frac{k}{\mu + k}\right)^k.
\end{equation}
The unconditional probability of observing a number of infestations is a mixture:
\begin{equation} \label{eq:mix}
\text{Pr}(I_t(x)) = \begin{cases} 
(1-\pi_t(x)) + \pi_t(x) \cdot f(I_{t}(x)) & \text{if } I_t(x) = 0, \\
\pi_t(x) \cdot f(I_{t}(x)) & \text{if } I_t(x) > 0.
\end{cases}
\end{equation}


The calculation of beetle pressure relies on a dispersal submodel which maps last year's infestations to the number of arriving beetles after the summer dispersal phase. The \textit{beetle pressure} at cell \( y \) in year \( t \), denoted \( b_t(y) \), is derived by convolving the previous year's infestations with a dispersal probability function. For notational simplicity, we have defined the beetle pressure \textit{predictor} as $B = log(b)$. The beetle pressure \textit{per se} is defined as
\begin{equation} \label{eq:beetle_pressure}
b_t(y) = \sum_{x} I_{t-1}^*(x) \theta_t(x) \bar{D}\left(\text{dist}(y, x)\right),
\end{equation}
where $I_{t-1}^*(x)$ denotes uncontrolled infestations from the previous year, and $\theta_t(x)$ denotes the probability of winter survival, calculated with the model of \citet{regniere2007modeling} along with imputed weather data from \textit{BioSIM} \citep{regniere2014biosim}. The function $\bar{D}$ gives the probability of dispersal between cells. The product $I_{t-1}(x) \, \theta_t(x)$ is approximately proportional to surviving beetle progeny in cell $x$.  We use the word \textit{proportional} because the Heli-GPS and concentric surveys count infested trees, not beetles. 

The dispersal kernel is a radially-symmetric Student's t-distribution, with a scale parameter $\sigma$ and \textit{degrees of freedom parameter}, $\nu$. The student-t distribution interpolates between a Gaussian distribution (which has exponentially decaying, ``thin tails'') as $\nu \rightarrow \infty$, and a fat-tailed Cauchy distribution as $\nu \rightarrow 1$. The radially symmetric kernel density $D$ in 2D space is parameterized as a function of the Euclidean distance $r = \text{dist}(x,y)$ between the center points of two cells:

\begin{equation} \label{eq:kernel_density}
D(r) = \frac{(\nu -1) \left(\frac{r^2}{\nu  \rho ^2}+1\right)^{\frac{1}{2} (-\nu -1)}}{2 \pi  \nu  \rho ^2}.
\end{equation}

Based on a previous study of MPB dispersal in western Alberta, we set $\nu = 1.45$ and $\rho = 0.0118$ \citep{johnson2024stratified}. This makes the resulting distribution similar to the fat-failed Cauchy distribution, with a median dispersal distance around 50 meters, and a $95^{\text{th}}$ percentile dispersal distance around 5 km. To more accurately represent dispersal at a 5$\times$5 km resolution, we performed a convolution on higher-resolution grids (50x50 m cells) using the discretized kernel mass ($D(r) \times \left(0.05\right)^2$), assuming that focal infestations are uniformly distributed within the central 5$\times$5 km square. The results are then aggregated to a coarser spatial scale, giving the transition probability function $\bar{D}$ used in \eqref{eq:beetle_pressure}.

\subsection{Model justification} \label{Model justification}

\begin{figure}[H]
\centering
\includegraphics[scale = 1]{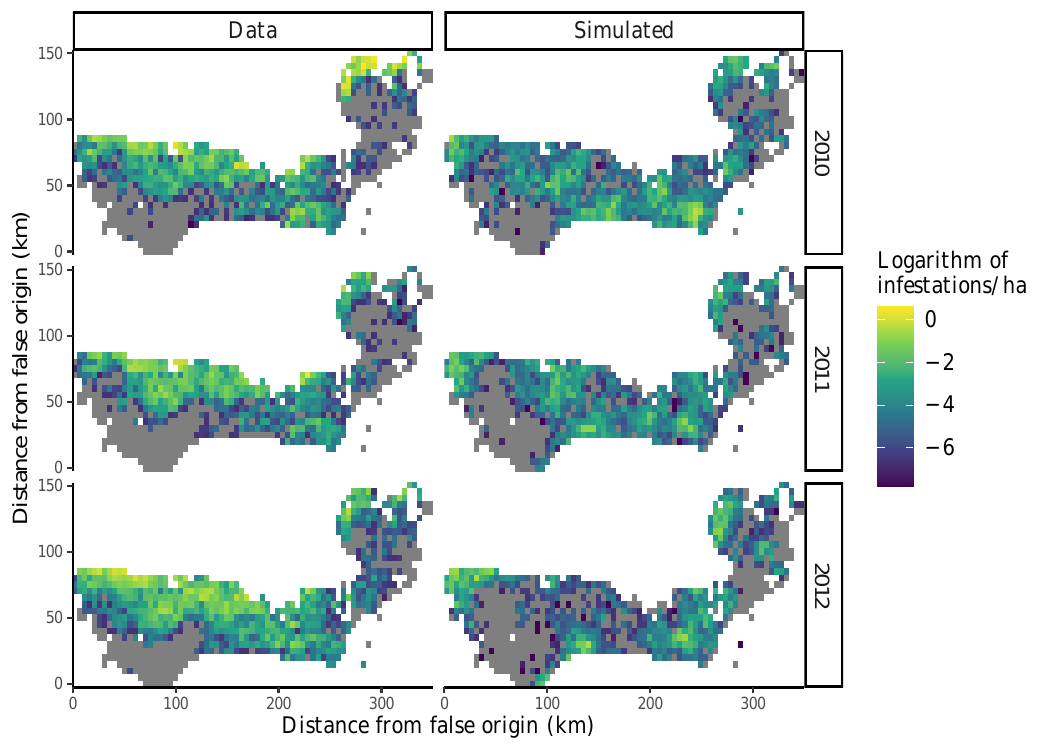}
\caption{The model recreates spatial patterns in the data, specifically the long-range spatial autocorrelation of infestations. Here the actual outbreak (left column) is compared to a single simulation for the years 2010--2013 (right column).}
\label{fig:sim_map}
\end{figure}

\begin{itemize}

\item Winter temperatures, pine volume/density, and current infestations are consistently found as the most important predictors of MPB dynamics \citep{aukema2008movement, ramazi2021predicting, srivastava2023dynamic}. This justifies our use of the overwintering survival values $\theta_t(x)$, as well as the two predictors $B$ and $V$. The cumulative killed trees predictor, $K$, was included in order to test the hypothesis that host-tree depletion contributed to the collapse of the outbreak. 

\item The logarithmic transformation involved in the calculation of all predictors ($B$, $V$, and $K$) was necessary to achieve the linearity assumptions, implicit in \eqref{eq:pi} \& \eqref{eq:mu} (evidence in Fig. \ref{fig:mod2_just}).

\item Pine volume was treated as constant over time. While pine volume can change due to logging, MPB damage, etc., we lacked data on forestry activities, and MPB infestations were not severe enough to significantly alter volumes. Alternatively, pine volume can be thought of as a proxy for the site index (a measure of potential forest productivity), which is approximately constant on the decadal time scale.

\item The ZINB model structure has previously been used to model MPB dynamics \citep{xie2024modelling, johnson2024mountain} and accurately captures the observed variation in infestation densities (Fig. \ref{fig:NB_pred_vs_obs}). The ZINB model with the 3 main predictors performs better than two alternative models, including the ZINB model with interaction effects, and a different, more mechanistic model (Appendix \ref{Model comparisons}).

\item The year effects were necessary because overwintering survival could not fully capture large year-to-year differences in infestation densities. There is not strong evidence for a relationship between year effects and overwintering survival (Fig. \ref{fig:Psurv_vs_c}).

\item The presence/absence and count submodels (\eqref{eq:pi} \& \eqref{eq:mu} respectively) share the same year effects with different scaling. Alternative models with independent year effects exhibited a strong correlation between the submodels' respective year effects (r = 0.87); a presentation of these models is omitted for brevity.

\item The Gaussian process was essential for removing residual autocorrelation, which can artificially narrow posterior distributions and lead to overconfident predictions about MPB control effectiveness. Models without the Gaussian processes produced an unrealistically narrow range of outcomes that failed to capture the observed trajectory of infestation densities (Fig. \ref{fig:time_series_ribbons_combined}). Models including Gaussian processes generated spatial patterns that visually match the real data (Fig. \ref{fig:sim_map}).

\end{itemize}

\subsection{Model fitting} \label{Model fitting}

All models were fit using \textit{Stan} \citep{stan2020rstan}, a Bayesian model-fitting program that implements Hamiltonian Monte Carlo. Gaussian processes code was provided by \citet{hoffmann2023scalable}. Standard model-fitting diagnostics were examined. Parameters were given weakly-informative parameters (in the sense of \citealp{gelman2014bayesian}) and the posterior contraction statistic  \citep{schad2021toward} showed that prior distributions generally had a small influence on parameter estimates (Appendix \ref{model_details}). 

\subsection{Simulations} \label{Simulations}

We conducted simulations across an extensive region in western Alberta (Fig. \ref{fig:study_areas}), using the 2009 infestations as the initial conditions. Recall that control efficiency $m$ is the proportion of infestations that are removed via level 1 treatment. If $I_{t-1}(x)$ now represents the most recent simulated values (or initial conditions), the beetle pressure is calculated as
\begin{equation} \label{eq:beetle_pressure2}
b_t(y) = m \, \sum_{x} I_{t-1}(x) \, c(x) \,  \theta_t(x) \bar{D}\left(\text{dist}(y, x)\right). 
\end{equation}
Our simulations use a spatiotemporally constant control efficacy, even though real-world efficacy varies with infestation density, proximity to untreated areas \citep{carroll2017assessing}, and management responses. We chose to keep $m$ constant given our imperfect understanding of fine-scale MPB dynamics and the proprietary nature of Alberta's decision support system. For with-control simulations, we set $m = 0.47$ (the median efficacy observed from 2009-2020 in consistently surveyed areas), and for no-control scenarios, $m = 0$.

We created additional counterfactual scenarios by manipulating parameters or predictors. To simulate the absence of severe winters, we replaced the cell and year-specific overwintering survival with a measure of central tendency, specifically the median of $\theta_t(x)$ across 2009-2020 in all cells where at least one infestation occurred over the same period. To remove host depletion effects, we set the killed-trees predictor at $K = 0$.

Our simulations fall into two distinct categories: \textit{actual outbreak} and \textit{future outbreak} scenarios; this categorization is independent of other simulation variations, such as the presence or absence of control measures. The actual outbreak scenarios use the specific environmental conditions observed from 2009-2019, specifically the overwintering survival lattices, $\theta_t(x)$, and year effects, $\eta_y$, in sequence. In contrast, future outbreak scenarios used randomly generated conditions. Temporally uncorrelated year effects are drawn from the standard normal distribution, and overwintering lattices are resampled from the collection of 2009--2019 lattices. This ``bootstrapped lattices'' approach maintains the spatial structure of $\theta_t(x)$ while introducing temporal randomness. 

For each scenario, we conducted 300 simulation runs, each using a different posterior sample of model parameters. Therefore, variation across simulation runs capture multiple sources of uncertainty: \textit{parameter uncertainty} from the posterior distribution, \textit{environmental stochasticity} from year effects and overwintering survival (but only in \textit{future outbreak scenarios}), and \textit{intrinsic stochasticity} arising from the Gaussian processes and negative binomial sampling.

\subsection{Management cost calculations}

To find the cost of level 1 treatment, we use arithmetic, previously published figures, and several assumptions. The cost of level 1 treatment, in units of Canadian dollars (CAD) per tree, is calculated by dividing the total cost by the total number of controlled trees from 2004-2016, using figures from \citet{hodge2017strategic}: 456,000,000 CAD / 1,430,000 trees = 318.9 CAD per tree. To calculate the total cost at any hypothetical control efficacy level, we multiply the cumulative number of controlled trees from 2009--2019 by 318.9 CAD. We use the years 2009--2019 because control in these years influences infestation densities in the years which we aim to predict: 2010--2020. For better interpretability, we can calculate the 11-year cost per hectare by dividing the total cost by the 2,110,000 ha in our study area.

Several assumptions underpin these calculations. We exclude the cost of heli-GPS surveys, but this is unproblematic because these costs are negligible (0.15 CAD per hectare; \citealp{wulder2006augmenting}) compared to the cost of ground surveys and tree removal. A second unproblematic assumption is the linear relationship between cost and number of controlled trees, which is implied by the linear relationship between cost and ground survey area \citep{kunegel2019management}; and between ground survey area and number of controlled trees (Fig. \ref{fig:area_vs_control}).

Two assumptions warrant more scrutiny. The first is the assumption of constant per-tree treatment costs, only holds true near the observed median control efficacy of 0.47 --- field crews are only sent to about 50\% of red-topped tree sites, so more green-attack trees could be found simply by mobilizing more field crews. However, achieving higher control efficacy becomes increasingly expensive on a per-infestation basis. Once crews have addressed most red-topped-tree sites, they must conduct wider concentric surveys at each site to attain higher efficacy. Expanding the survey radius from 50 m to 300 m increases efficacy from 55\% to 85\%, but at 36 times the cost \citep{kunegel2019management}. The second questionable assumption is treating control efficacy as constant when it actually fluctuates. Given these limitations, our analysis serves best as a framework for order-of-magnitude cost estimates. Given the highly stochastic nature of MPB outbreaks, precise quantitative predictions are not a reasonable goal in any case.

\section{Results} \label{Results}

\subsection{Variable importance}

Before delving into the simulation results, we examine the parameter estimates (Table \ref{tab:pars}) to identify the most important predictors and sources of variability. Beetle pressure is the most important predictor, followed by pine volume. The cumulative killed trees predictor has a minuscule coefficient, implying that host-tree depletion --- via MPB specifically, not logging --- is unlikely to explain any feature of MPB's dynamics over the study period. Both year effects and the Gaussian process are scaled by parameters with roughly similar magnitudes. Moreover, the sum of these parameters is similar to the effect of beetle pressure (e.g., compare $\tau_{\gamma} + \sigma_{\gamma}$ to $\gamma_B$), thus highlighting the importance of year effects and Gaussian processes in accounting for otherwise unexplained variation. Interestingly, the length scale parameter is approximately 10 km, which implies that residual autocorrelations are substantial (i.e., greater than 0.05) for distances up to 25 km.

\begin{table}[H]
\centering
\begin{tabular}{llllll}
  \hline
Parameter & Short description & Mean & SD & $\text{CI}_{2.5\%}$ & $\text{CI}_{97.5\%}$ \\ 
  \hline
$\gamma_{0}$ & Intercept, presence & 3.1 & 0.44 & 2.1 & 3.9 \\ 
  $\gamma_{B}$ & Beetle pressure, presence & 3.1 & 0.11 & 2.9 & 3.3 \\ 
  $\gamma_{V}$ & Pine volume, presence & 0.29 & 0.057 & 0.19 & 0.41 \\ 
  $\gamma_{K}$ & Trees killed, presence & -0.0076 & 0.0073 & -0.026 & -0.00022 \\ 
  $\tau_{\gamma}$ & Year effect scale, presence & 1.3 & 0.36 & 0.77 & 2.2 \\ 
  $\sigma_{\gamma}$ & GP scale, presence & 1.9 & 0.10 & 1.7 & 2.1 \\ 
  $\beta_{0}$ & Intercept, count & 3.4 & 0.16 & 3.0 & 3.7 \\ 
  $\beta_{B}$ & Beetle pressure, count & 1.6 & 0.024 & 1.5 & 1.6 \\ 
  $\beta_{V}$ & Pine volume, count & 0.15 & 0.013 & 0.13 & 0.18 \\ 
  $\beta_{K}$ & Trees killed, count & -0.0021 & 0.0021 & -0.0077 & -0.000077 \\ 
  $\tau_{\beta}$ & Year effect scale, count & 0.45 & 0.14 & 0.25 & 0.76 \\ 
  $\sigma_{\beta}$ & GP scale, count & 1.0 & 0.029 & 0.95 & 1.1 \\ 
  $L$ & GP length scale & 11 & 0.34 & 10 & 11 \\ 
  $k$ & Dispersion parameter, count & 3.2 & 0.096 & 3.0 & 3.4 \\ 
   \hline
\end{tabular}
\caption{Parameter estimates from the main model. The coefficients here apply to the \textit{standardized} versions of the predictors ($B$, $V$, and $K$) so that the coefficient magnitudes can be interpreted as \textit{predictor importance}. The metrics $\text{CI}_{2.5\%}$ \& $\text{CI}_{97.5\%}$ are the bounds of the 95\% credible intervals.} 
\label{tab:pars}
\end{table}

\subsection{What ended the outbreak?} 

\begin{figure}[H]
\centering
\includegraphics[scale = 1]{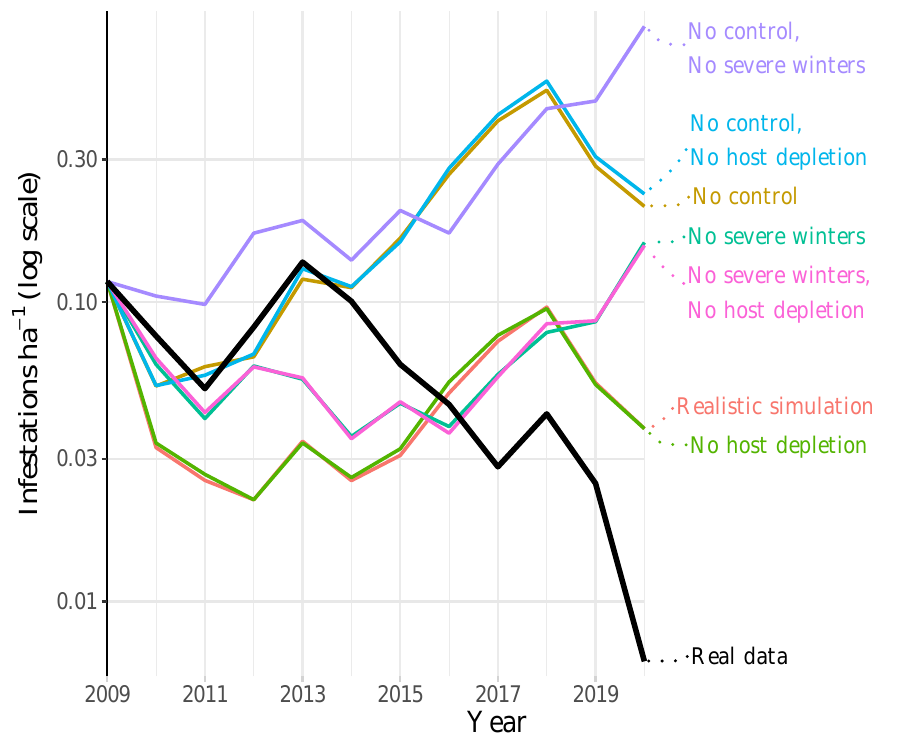}
\caption{Time series of mountain pine beetle infestation densities, in both counterfactual simulation scenarios (colored lines show the median value across simulations), and in the real data (black line). Simulations use the observed environmental parameters from 2009--2019, following the \textit{actual outbreak} approach (see Section \ref{Simulations} for details). The y-axis values represent the spatial average of infestations $\text{ha}^{-1}$ across the study area.}
\label{fig:typical_TS_main}
\end{figure}

A combination of control efforts and a sequence of severe winters ended the outbreak (Figure \ref{fig:typical_TS_main}). When we simulated scenarios removing either control efforts or severe winters individually, the resulting infestation densities were approximately ten times higher than what was actually observed. Even more striking, simulations that removed both factors predicted infestation densities in 2020 that were approximately 100 times higher than observed levels. Host tree depletion, as expected, had no appreciable effect on outbreak trajectories.

Readers examining Figure \ref{fig:typical_TS_main} may notice that none of the counterfactual simulations achieve the low infestation density shown in the ``Real data'' line. This apparent discrepancy has two explanations. First, the counterfactual trajectories represent median values across multiple simulations, while the real data represents a single instance of a stochastic process. Second, the healthy pine strategy, which is not included in our model due to data limitations, may have contributed to further reducing infestation densities.

\subsection{What factors reduced the cumulative number of infestations?}

\begin{figure}[H]
\centering
\includegraphics[scale = 1]{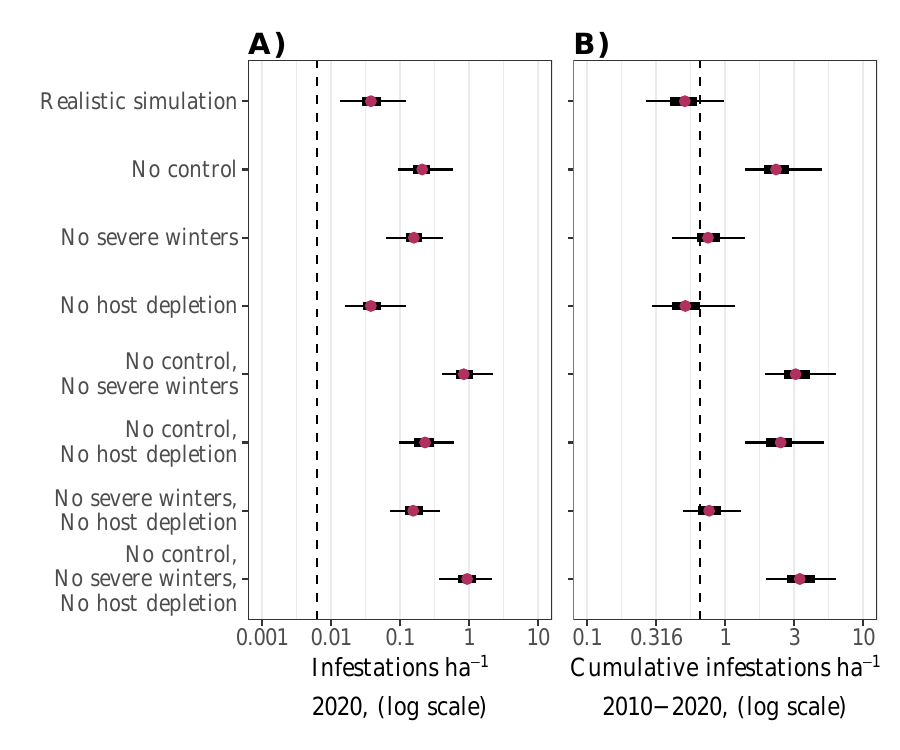}
\caption{Outbreak quantities of interest (x-axis) across various simulation scenarios (y-axis). Points, horizontal thick lines, and thin lines respectively show the mean, 50\% interval, and 95\% interval of the predictive posterior distributions. All values are spatial averages across the study area. Dashed vertical lines show the actual value. Simulations use the observed environmental parameters from 2009--2019, following the \textit{actual outbreak} approach (see Section \ref{Simulations} for details).}
\label{fig:boxplots_main}
\end{figure}

Level 1 treatment is the primary factor limiting total infestations over the study period (Fig. \ref{fig:boxplots_main}, Panel B). Although severe winters contributed to ending the outbreak, their impact on the cumulative number of infested trees was minimal, since most infestations occurred between 2009 and 2018 before the onset of unusually cold winters.


\subsection{How many trees were saved by control efforts?} 

Over the 2009-2020 period, control efforts prevented the deaths of 1.8 trees per hectare, with a 95\% posterior predictive interval (95\% PI) of 0.91--4.1 trees per hectare (Table \ref{tab:control_delta1}). This amounts to a 79\% (95\% PI:58--89\%) reduction in tree death compared to a do-nothing scenario. Multiplied over the 2,110,000 total hectares of the consistently surveyed area (i.e., the black polygon in Figure \ref{fig:study_areas}), this translates to 3,798,000 trees saved over the 11-year period. Control effectiveness can also be expressed as the \textit{control multiplier} which shows that removing one tree through level 1 treatment prevents approximately 6.7 (95\% PI:2.6--15) infestations in the long run.

The \textit{actual outbreak} simulations, which incorporate observed environmental conditions and level 1 treatment, produced a 95\% predictive interval for cumulative infestations of 0.26--0.98 trees per hectare, with the observed value of 0.66 trees per hectare falling near the center of this range. This agreement between model predictions and actual outcomes is noteworthy, as it demonstrates the model's ability to accurately predict long-term system behavior despite being parameterized using 1-year-ahead relationships. In simulations exploring both the \textit{no control} and \textit{future outbreak} scenarios, the upper estimate of cumulative infestations is 8.0 trees per hectare. This figure is also reasonable based on our experience with aerial overview survey data from the western United States. While infestation densities were an order of magnitude higher during the 2000s outbreak in British Columbia, this can be attributed to the dense monoculture of mature lodgepole pine in central BC, a feature not replicated in Alberta's sparser and more mixed forests.

\renewcommand*\rot{\multicolumn{1}{R{30}{1em}}}
\renewcommand\cellset{\renewcommand\arraystretch{0.8}%
\setlength\extrarowheight{0pt}}
\begin{table}[H]
\begingroup\fontsize{9pt}{10pt}\selectfont
\begin{tabular}{lllllll}
\rot{Simulation scenario}  & \rot{Trees Killed $\text{ha}^{-1}$ (No control)} & \rot{Trees Killed $\text{ha}^{-1}$ (Control)} & \rot{No. trees $\text{ha}^{-1}$ Saved by control} & \rot{\% Trees Saved by control}& \rot{Control multiplier} \\ 
  \hline \\
\makecell[tr]{Actual outbreak,\\2010--2020} & 2.3 (1.4-5) & 0.51 (0.26-0.98) & 1.8 (0.91-4.1) & 79 (58-89) & 6.7 (2.6-15) \\ 
  \makecell[tr]{Future outbreak,\\11 year period} & 2.2 (0.41-9.7) & 0.55 (0.089-2.4) & 1.5 (0.26-8) & 74 (43-88) & 4.9 (1.1-13) \\
   \hline
\end{tabular}
\endgroup
\caption{How many trees did control efforts save over an 11-year period? \textit{No control} vs. \textit{Control} columns correspond to a control efficacy of $m= 0$ vs. $m=0.47$ control; units are trees per hectare, averaged across the study area. The format of table entries is \textit{Median (95\% predictive intervals)}. The \textit{actual outbreak} simulation scenario uses estimated values of overwintering survival and year effects. The \textit{Future outbreak} scenario uses bootstrapped maps of overwintering survival and randomly generated year effects. The \textit{control multiplier} is calculated as the number of trees saved by control from 2010--2020, divided by the number of controlled trees from 2009--2019.} 
\label{tab:control_delta1}
\end{table}

\subsection{What range of treatment outcomes should be expected in future outbreaks?} 

In the future where environmental conditions are unknown, outbreak trajectories are highly variable (second row in Table \ref{tab:control_delta1}). Under a no-control scenario, MPB may kill anywhere from 0.41 to 9.7 trees per hectare. Under the control scenario, MPB may kill anywhere from 0.089 to 2.4 trees per hectare. Importantly, the difference between the control and no-control scenarios under the observed environmental conditions (i.e., $2.3-0.51 \approx 1.8$) is smaller than the range of outcomes under a control scenario but unknown environmental conditions (i.e., $2.4-0.09 \approx 2.3$). Put another way, the effect of environmental stochasticity is comparable to the effect of control. This comparison suggests that future outbreaks could potentially result in significantly higher infestation densities even with control measures in place. However, such outcomes should not be interpreted as a failure of control methods. The relative effectiveness of control in the uncertain \textit{future outbreak} scenario is similar to the relative effectiveness in the \textit{actual oubreak} scenario (43-88\% and 58--89\% respectively).

\subsection{How does MPB control cost vary with control intensity?}

\begin{figure}[H]
\centering
\includegraphics[scale = 0.8]{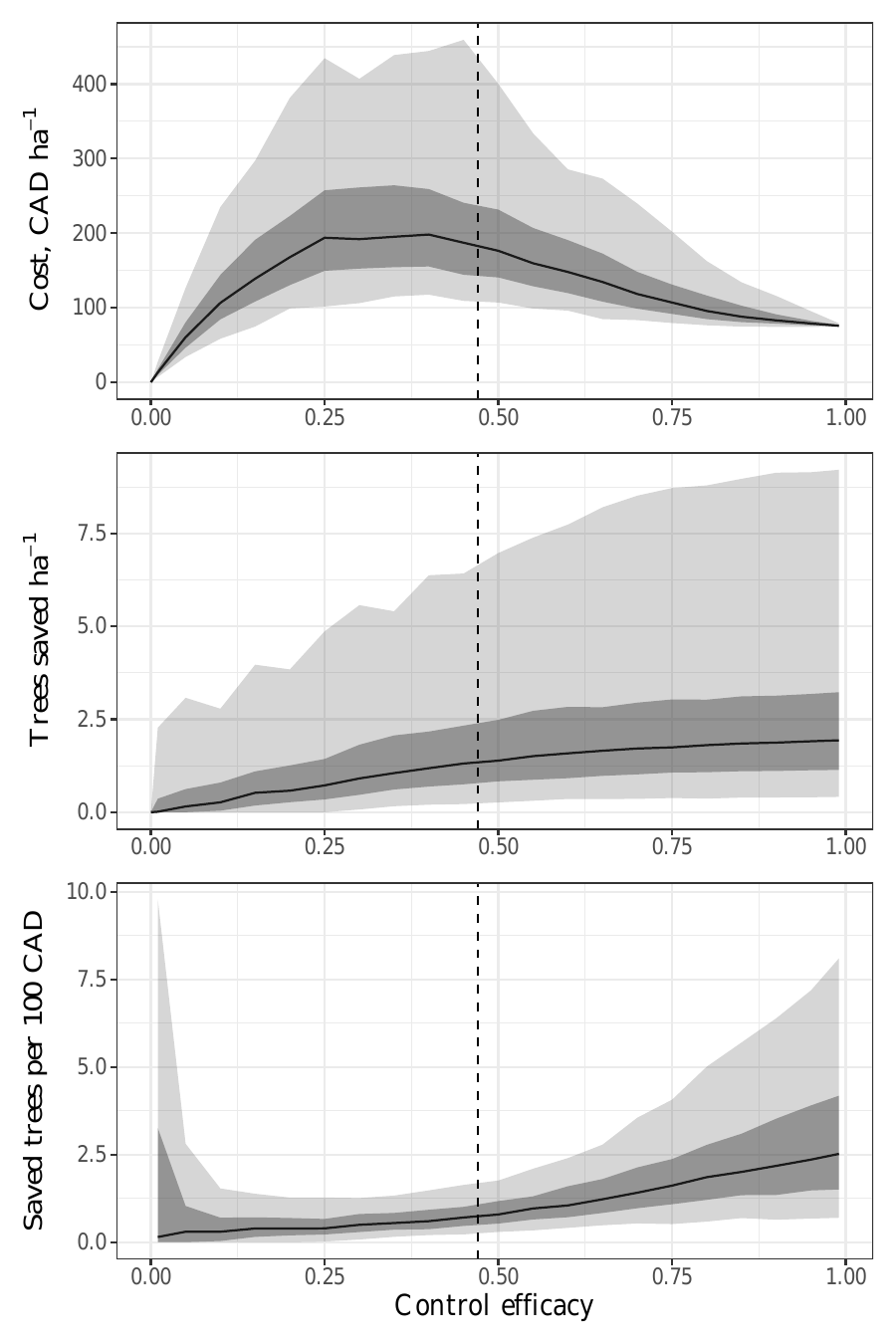}
\caption{Cost, effectiveness (trees saved), and efficiency (trees saved per 100 CAD), as a function of control efficacy (proportion of infestations removed). Lines and grey ribbons show the median, 50\%, and 95\% predictive intervals across \textit{future outbreak} simulations. The vertical dashed line shows the actual control efficacy (median across 2009--2019). Y-axis values show the cumulative number of trees over an 11-year period (2009--2019 for controlled trees, 2010--2020 for saved trees), averaged over the study area. The actual cost was approximately 110 CAD per hectare for the 2009--2019 time period.}
\label{fig:combined_costs}
\end{figure}

Figure \ref{fig:combined_costs} illustrates three key metrics of mountain pine beetle (MPB) control: cost (CAD $\text{ha}^{-1}$), effectiveness (trees saved $\text{ha}^{-1}$), and efficiency (saved trees per 100 CAD), across varying levels of control efficacy (proportion of all infestations controlled with level 1 treatment). For the sake of clarity, we explicitly distinguish between the meanings of similar-sounding ``E'' words: efficacy measures immediate or controlled-setting results (operationalized here as the proportion of infested trees removed via level 1 treatment), effectiveness evaluates long-term impact, and efficiency is the quotient of outcomes to costs.

The relationship between control efficacy and cost is hump-shaped. Near the median observed value of  47\% control efficacy (vertical line in Fig. \ref{fig:combined_costs}), costs range from 100--400 CAD per hectare, with this wide range reflecting uncertainty in future outbreak trajectories. More explosive outbreaks require treating more infested trees, increasing costs. Interestingly, the highest costs occur at intermediate control efficacy levels, where beetle populations continue growing despite control efforts. Both low and high control efficacies show lower costs, but for different reasons: low efficacy requires minimal intervention and is therefore inexpensive, while high efficacy completely suppresses outbreaks, leading to fewer treatments in the long run.

\begin{figure}[H]
\centering
\includegraphics[scale = 1]{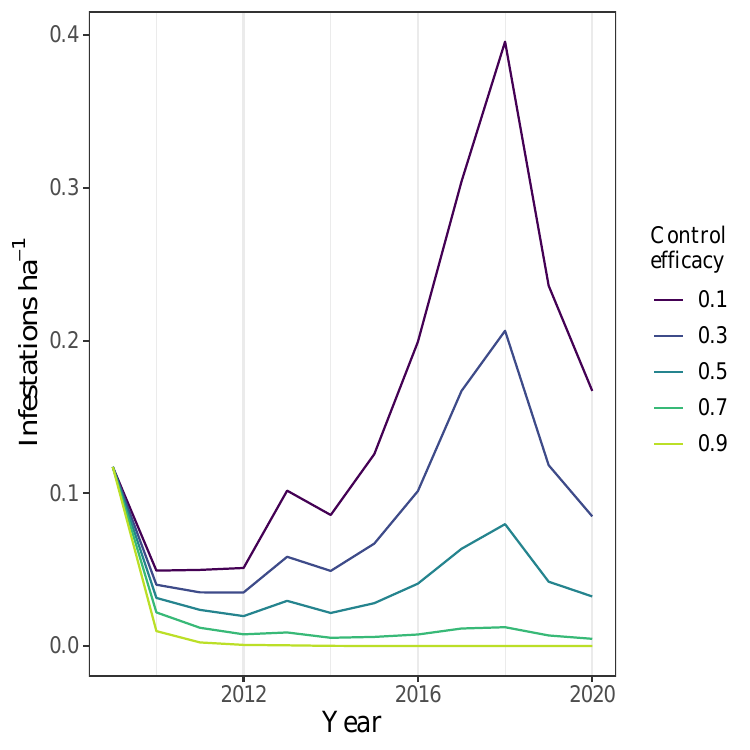}
\caption{Infestation densities increase exponentially as control efficacy decreases. Lines show the median value across simulations, of the infestations $\text{ha}^{-1}$ averaged across the study area. Simulations use the observed environmental parameters from 2009--2019, following the \textit{actual outbreak} approach (see Section \ref{Simulations} for details).}
\label{fig:TS_control_sfit2_gp_collapse_5000m}
\end{figure}

Increasing control efficacy increases efficiency (measured as trees saved per 100 CAD), with higher efficacy levels yielding disproportionate benefits. This not only supports \citepos{carroll2006direct} conclusion that early and intense control is most effective, but extends it by showing that such a strategy is also more efficient/cost-effective. The relationship between efficacy and efficiency is notably nonlinear, driven by the geometric growth of unchecked MPB populations. This is illustrated in Figure \ref{fig:TS_control_sfit2_gp_collapse_5000m}, where small decreases in control efficacy lead to exponential increases in infestation densities.

\section{Discussion} \label{Discussion}

Our analysis revealed several insights about the long-term effectiveness of Alberta's control strategy. Level 1 treatment reduced the total number of infestations by 79\% ($95\%$: 82--89\%) over an 11-year period, and was the only factor that substantially reduced total damage. Neither cold winters nor host-tree depletion (specifically from MPB, not logging) had a significant impact. However, cold winters and level 1 treatment were equally important in causing population collapse in the late 2010s. The trajectory of future outbreaks and control-effectiveness is highly uncertain due to stochastic environmental conditions. This uncertainty also means that achieving the same control efficacy (approximately 50\%) in future outbreaks could cost anywhere from 100 to 400 CAD per hectare. While higher control efficacy (the proportion of all infestations that are treated) would reduce costs by completely suppressing outbreaks, achieving high efficacy likely requires technological advances, in particular the remote sensing of early-infested green-attack trees.

Cold winters and Level 1 treatment acted synergistically in terminating the outbreak, lending some support to a ``wait it out'' strategy, wherein moderate control effort is applied until extreme cold weather occurs. MPB population crashes due to extreme cold have been documented since the 1930s \citep{evenden1940destructive}, but only two instances of collapse at a provincial/state scale have been recorded. The first was in British Columbia, where an emerging epidemic was stopped by a single winter in 1984--1985, with a cold snap in late October \textit{and} minimum temperatures reaching -43C \citep{safranyik1991unseasonably, stahl2006climatology}. This paper establishes the Alberta collapse as a second example, with the notable distinction that the collapse occurred over a sequence of cold winters. Our analysis likely understates the impact of cold temperatures, since it only includes data through the winter of 2019 (beginning in December 2019), truncating the series of unusually cold winters that occurred from 2018 to 2023.

The trajectories of future MPB outbreaks are highly uncertain. This partially stems from fluctuating environmental factors (here overwintering mortality and year effects), with our main model showing that the cumulative killed trees may vary across an order of magnitude. Similarly, the absolute effectiveness of control --- measured as the number of trees saved over an 11-year period --- varies across an order of magnitude. Another source of uncertainty is model uncertainty, as demonstrated by an alternative model (Appendix \ref{Alternative models}) where uncontrolled outbreaks are 5 times more destructive than in the main model. Further uncertainty stems from factors that are extremely difficult to model, including the initial conditions of future outbreaks, long-distance dispersal from BC, and the uncertain interaction between climate change and MPB dynamics \citep{cooke2017predicting, bentz2022complexities, brush2023coupling}. Readers should focus on orders of magnitude or relative metrics (i.e., proportion of trees saved) as these better reflect the limitations of forecasting populations with highly stochastic dynamics.

Metrics of control effectiveness like ``trees saved'' might seem shortsighted given  Canada's extensive, self-regenerating forests, but they serve as proxies for how control may mitigate the broader impacts of MPB outbreaks. Mountain pine beetle outbreaks cause economic harm \citep{abbott2009mountain, corbett2016economic, price2010insect, rosenberger2013estimating} and have many ecological consequences, including shifts in forest structure from even-aged to uneven-aged stands, improved habitat for woodpeckers and saproxylic beetles, and altered carbon dynamics (\citealp{dhar2016aftermath}, and sources therein). Indeed, tree decomposition from the severe 2000s outbreak converted BC forests from a carbon sink to a carbon source for at least 15 years \citep{kurz2008mountain}. Some ecological consequences may be viewed as beneficial, such as an increase in the diversity of understory plants and certain mammals following an outbreak \citep{saab2014ecological, pec2015rapid}. Last but not least, MPB outbreaks alter hydrological functioning in a variety of ways \citep{schnorbus2011synthesis, redding2008mountain}.

We explicitly do not make recommendations about whether control measures were worthwhile or how much should be invested in future control efforts. Instead, we aim to provide information to support public discourse and decision-making. Forest managers and government officials are better positioned to make complex ecological management decisions, as they possess more contextual information, including funding availability, logistical limitations, stakeholder requirements, and political factors. The broader public --- as taxpayers and stakeholders --- also has a legitimate interest in understanding and engaging with these ecological management decisions. 

Our study builds on previous control assessments in two ways. First, we evaluate multiple sources of uncertainty, including environmental stochasticity (i.e., climatic fluctuations), parameter uncertainty, and model uncertainty. Second, our model is simpler than previous models in order to balance realism with additional goals: avoiding overfitting, avoiding manual parameter tuning, and justifying model structure (Section \ref{Model justification}) with the hope of reducing model bias. Despite substantially different modeling approaches, our results align with those of \citet{carroll2017assessing}: we find that control measures save 58--89\% of trees, comparable to their finding of approximately 66\% saved pine volume. This agreement can be verified by examining Figure 9 in \citet{carroll2017assessing}, specifically comparing the StS'' and StS ends 2020'' scenarios from the 2020-2030 time period.

Two model limitations warrant reiteration. Simulations do not incorporate Alberta's healthy pine strategy (i.e., removing susceptible stands with preemptive clearcutting or prescribed burns), because the necessary data was unavailable. This omission may explain why our model overpredicted infestation densities in 2020 (Fig. \ref{fig:typical_TS_main}). Additionally, the model assumes uniform control efficacy across the landscape, whereas actual control efforts vary across space and time. This simplification was necessary due to our limited understanding of the actual decision process for allocating control effort. An important consequence is that we likely underestimate the effectiveness of control, as strategic targeting would presumably be more effective than constant control effort.

What comes next for MPB in Alberta?  Since MPB affected only a small fraction of lodgepole pine, outbreaks could resurge within decades, though this depends on the yet-to-be-assessed effectiveness of Alberta's healthy pine strategy. Unlike British Columbia's widespread, episodic outbreaks that resulted from fire suppression and extensive lodgepole pine monocultures, Alberta's mixed forests may experience smaller, asynchronous outbreaks similar to those observed in the western United States from 1950-2000 \citep{axelson2018stand}. While climate change and warming temperatures are expected to increase the frequency of outbreaks \citep{brush2023coupling}, low-elevation MPB populations might suffer from accelerated development that disrupts the timing of their life-cycle (i.e., fractional voltinism; \citealp{bentz2022complexities}). The beetle's eastward expansion through Alberta critically depends on its adaptation to jack pine, a novel host tree. Experimental evidence suggests jack pine is suitable for MPB development \citep{safranyik1982survival, cerezke1995egg, rosenberger2017cold, musso2023pine}, but MPB may struggle to detect and coordinate mass attacks due to the jack pine's smaller size and distinct chemical profile \citep{johnson2024mountain}. Additionally, competition from wood-boring beetles in jack pine forests \citep{pokorny2021novel} may prevent MPB from establishing endemic populations in eastern Alberta.

Consistent control efforts reduced total tree mortality by roughly three-quarters over eleven years, demonstrating humans' capacity to meaningfully influence forest pest dynamics. The Alberta case also reveals limitations. Achieving complete outbreak suppression requires both technological advances --- the remote-sensing of early-infested green attack trees --- and sustained funding commitments, which may become increasingly difficult to secure amid the rise of anti-science political movements in the United States and Canada. The cost-effectiveness of MPB control (trees saved per dollar) increases as more money is spent, but there are no guarantees. Even with simplifying assumptions (e.g., no dispersal from BC, no climate change effects), the trajectory of the next outbreak is highly uncertain, as is the absolute effectiveness of control. The broader question of whether the ecological benefits of MPB control justify the economic costs extends beyond the scope of this paper, but our study offers a powerful lesson: human intervention in complex ecological systems holds both great potential and great uncertainty.

\section{Acknowledgements} 

The authors would like to thank Allan Carroll, Barry Cooke, Kathy Bleiker, Antonia Musso, Micah Brush, and K\'evan Rastello, and Xiaoqi Xie for helpful discussions. Funding for this research has been provided through grants to the TRIA-FoR Project to ML from Genome Canada (Project No. 18202) and the Government of Alberta through Genome Alberta (Grant No.~L20TF), with contributions from the University of Alberta.

\section{Credits}
EJ conceived of the study, performed the analysis, and wrote the first draft; All authors contributed critically to the drafts and gave final approval for publication.

\section{Data availability statement}


Code and data will be made publicly available on Zenodo (\url{TBD}). 



\newpage

\begin{appendices}
\counterwithin{figure}{section}
\counterwithin{table}{section}

\section{Alternative models} \label{Alternative models}

To examine how model structure influences inferences, we analyze three different models.  The first is called the \textit{mechanistic model}. This model, inspired by recent work (\citealp{johnson2024explaining}), encodes geometric growth with per capita growth rates that decrease when there are not enough susceptible trees. Because the initial density of susceptible trees is unknown, we assume it is proportional to pine volume. The second model is the model presented in the main text. We call this the \textit{phenomenological model} because we used graphical evidence to determine the relationships between predictors and response variables, e.g, log(infestations) is a linear function of log(beetle pressure). The third model is the \textit{phenomenological model with interaction effects}. All these models blend mechanistic and phenomenological elements, but we use these labels to mark their relative positions on the phenomenological-to-mechanistic spectrum.

Using cross-validation and a number of predictive metrics, we show that these two new models --- the mechanistic model and the phenomenological model with interaction effects --- make worse predictions than the main phenomenological model. In absolute terms, however, all models perform well over short time horizons. For example, all models achieve an accuracy of 0.8--0.84 when predicting the presence or absence of infestations. 

Despite making similar 1-year predictions, the three models vary significantly in their long-term predictions, especially under counterfactual scenarios (e.g., no control). Much higher infestation densities can be attained under the mechanistic model. This makes sense because population growth in the mechanistic model is assumed to be proportional to population size i.e., $I_{t+1}(x) \propto c_t \times b_t(x)$. The mechanistic model, on the other hand, has an intercept parameter that creates a sublinear relationship between infestations and beetle pressure. 

The analysis of alternative models suggests a general skepticism of precise estimates in the context of MPB forecasting. In the face of huge model uncertainty, one ought to rely on qualitative results (e.g., more control is more efficient) relative metrics (e.g., percentage of trees saved by control), or order-of-magnitude estimate (e.g., without control, future outbreaks will kill 1s or 10s of trees per hectare over an 11-year period).

\subsection{Mechanistic model descriptions} \label{Mechanistic model descriptions}

The survival probability during the dispersal phase, denoted $s_t$, increases logistically with the density of susceptible trees:
\begin{equation}
    s_t(x) = \left(1 + \exp\left(-\lambda_1 \left( S_t(x) - \lambda_0\right) \right) \right)^{-1}.
\end{equation}
In this equation, $S_t$ represents the density of susceptible host trees in year $t$ (measured in trees per hectare). The parameter $\lambda_0$ serves as a threshold value, below which the survival probability is less than 0.5. The parameter $\lambda_1$ determines how quickly survival probability transitions between low and high values.

The initial density of susceptible trees is calculated as a proportion of pine volume (units: $\text{m}^3 \text{ha}^{-1}$):
\begin{equation}
    S_{0}(x) = (\text{pine volume}) \times \alpha.
\end{equation}
For subsequent years ($t > 0$), we calculate susceptible tree density by subtracting all previously infested trees:

\begin{equation}
    S_{t}(x) = \text{Max} \left(0,\,\, S_{0}(x) - \sum_{j=0}^{t-1} I_{j}(x)\right).
\end{equation}

The density of attacking beetles after dispersal-phase mortality is proportional to the intermediate quantity $A_{t}(x)$, which is defined as
\begin{equation}
   A_{t}(x)= c_t \, b_t(x) \, s_t(x) + \sigma_{A} f_t(x).
\end{equation}
Here, $b_t(x)$ represents beetle pressure, which is calculated identically to beetle pressure in the main model (see \eqref{eq:beetle_pressure}). The $c_t$ are year-specific productivity parameters that are analogous to the year effects in the main model. The $f_t(x)$ is a realization of a Gaussian process with length scale $L$ and scale $\sigma_A$. The covariance function is the same as that in the main model (see \eqref{eq:cov}).

The productivity parameters are gamma-distributed:
\begin{equation}
    c_t \sim \text{Gamma}\left(mean = \mu_c, sd = \sigma_c \right).
\end{equation}
Note that the gamma distribution is parameterized using mean and standard deviation so that hyperparameter estimates are more interpretable (Table \ref{tab:pars_mech}). The hierarchical structure enables us to simulate random productivity parameters for \textit{future outbreak} scenarios.

The probability of observing one or more infestations in a cell is denoted $\pi_t(x)$. This probability is a logistic function of $\log(A_t(x))$:

\begin{equation}
\pi_t(x) = \text{logit}^{-1}\left(\gamma_0 + \gamma_1 \log(A_t(x)) \right).
\end{equation}

Conditioned on the presence of one or more infestations, the probability density is  $f(I_{t}(x)) = \text{Normal}\left(\log(I_{t}(x)) | \mu_t(x), \sigma\right)$, where $\mu_t(x) = \log(A_t(x))$.

The complete probability density function is a statistical hurdle model: 
\begin{equation}
\text{PDF}(I_t(x)) = \begin{cases} 
1 - \pi_t(x) & \text{if } I_t(x) = 0, \\
\pi_t(x) \cdot f(I_{t}(x)) & \text{if } I_t(x) > 0.
\end{cases}
\end{equation}

There are several small differences between the mechanistic model and the phenomenological model.. First, the mechanistic model treats $I_t(x)$ as a continuous rather than a discrete quantity, calculating the probability density rather than the probability mass. This is because $I_x(t)$ represents density values (infestations per hectare) rather than integer counts (infestations per cell). We chose density measurements both to determine prior scales and because the log-linear hurdle model had fewer computational problems during model development. Second, the models generate zero-values differently. The mechanistic model is a hurdle framework, whereas the main text is a zero-inflated model. In a zero-inflated model, zero values can arise from either the presence-absence submodel or the count submodel, whereas a hurdle model only generates zeros through its presence-absence submodel.

\begin{table}[H]
\centering
\begin{tabular}{llllll}
  \hline
Parameter & Short description & Mean & SD & $\text{CI}_{2.5\%}$ & $\text{CI}_{97.5\%}$ \\ 
  \hline
$\alpha$ & Susc. trees per $\text{m}^3$ pine volume & 0.21 & 0.061 & 0.11 & 0.35 \\ 
  $\lambda_{0}$ & Susc. tree density threshold
for MPB survival & 4.4 & 4.1 & 0.12 & 15 \\ 
  $\lambda_{1}$ & Dispersal phase survival
transition rate & 0.046 & 0.017 & 0.022 & 0.086 \\ 
  $\gamma_{0}$ & Intercept, presence & 12 & 0.36 & 11 & 12 \\ 
  $\gamma_{1}$ & Attacking beetle effect, presence & 1.8 & 0.057 & 1.7 & 2.0 \\ 
  $\mu_{c}$ & Productivity mean & 9.1 & 1.7 & 6.2 & 13 \\ 
  $\sigma_{c}$ & Productivity std. dev. & 5.9 & 1.5 & 3.6 & 9.4 \\ 
  $\sigma_{A}$ & GP scale & 1.1 & 0.028 & 1.0 & 1.1 \\ 
  $L$ & GP length scale & 9.8 & 0.29 & 9.2 & 10 \\ 
  $\sigma$ & Residual std. dev. & 0.68 & 0.0092 & 0.66 & 0.70 \\ 
   \hline
\end{tabular}
\caption{Parameter estimates from the mechanistic model. ``Susc.'' is an abbreviation of susceptible. The metrics $\text{CI}_{2.5\%}$ \& $\text{CI}_{97.5\%}$ are the bounds of the 95\% credible intervals.}
\label{tab:pars_mech}
\end{table}

\subsection{Phenomenological model with interaction effects}

This model is just like the phenomenological model in the main text, except new terms are added to represent the interaction between predictor variables. The probability of potentially observing a non-zero number of infestations, previously \eqref{eq:pi}, becomes

\begin{align} \label{eq:pi2}
\pi_t(x) = \text{logit}^{-1}(
    &\gamma_0 + \gamma_B B_t(x) + \gamma_V V(x) + \gamma_K K_t(x)  \nonumber \\
     + & \gamma_{BV} B_t(x) \cdot V_t(x) + \gamma_{BK} B_t(x) \cdot K_t(x)  \gamma_{VK} V_t(x) \cdot K_t(x)  \nonumber \\
    + &\tau_{\gamma} \eta_t + \sigma_{\gamma} f_t(x)).
\end{align}

Similarly, the count submodel mean, previously \eqref{eq:mu}, becomes
\begin{align} \label{eq:mu2}
\mu_t(x) = \exp[
    &\beta_0 + \beta_B B_t(x) + \beta_V V(x) + \beta_K K_t(x)  \nonumber \\
   + &\beta_{BV} B_t(x) \cdot V_t(x) + \beta_{BK} B_t(x) \cdot K_t(x) \beta_{VK} V_t(x) \cdot K_t(x)  \nonumber \\
    + &\tau_{\beta} \eta_t + \sigma_{\beta} f_t(x)].
\end{align}

The parameters $\gamma_{BK}$ and $\beta_{BK}$ are constrained to be negative, in order to prevent run-away explosive population growth. We attempted to fit an even more complex model with quadratic effects (with terms like $ \gamma_{BB} B_t(x)^2$), but we encountered computational difficulties (divergent trajectories in Hamiltonian Monte Carlo). Parameter estimates are given in Table \ref{tab:pars_int}.

\begin{table}[H]
\centering
\begin{tabular}{llllll}
  \hline
Parameter & Short description & Mean & SD & $\text{CI}_{2.5\%}$ & $\text{CI}_{97.5\%}$ \\ 
  \hline
$\gamma_{0}$ & Intercept, presence & 3.2 & 0.38 & 2.5 & 4.0 \\ 
  $\gamma_{B}$ & Beetle pressure, presence & 2.9 & 0.11 & 2.7 & 3.1 \\ 
  $\gamma_{V}$ & Pine volume, presence & 0.29 & 0.069 & 0.15 & 0.43 \\ 
  $\gamma_{K}$ & Trees killed, presence & -0.018 & 0.017 & -0.060 & -0.00047 \\ 
  $\gamma_{BV}$ & Beetle pressure $\times$ pine volume, presence & 0.13 & 0.069 & -0.013 & 0.26 \\ 
  $\gamma_{BK}$ & Beetle pressure $\times$ trees killed, presence & -0.45 & 0.063 & -0.58 & -0.34 \\ 
  $\gamma_{VK}$ & Volume $\times$ trees killed, presence & -0.10 & 0.054 & -0.21 & 0.0019 \\ 
  $\tau_{\gamma}$ & Year effect scale, presence & 1.2 & 0.35 & 0.65 & 1.9 \\ 
  $\sigma_{\gamma}$ & GP scale, presence & 1.9 & 0.095 & 1.7 & 2.1 \\ 
  $\beta_{0}$ & Intercept, count & 3.4 & 0.16 & 3.1 & 3.7 \\ 
  $\beta_{B}$ & Beetle pressure, count & 1.6 & 0.023 & 1.5 & 1.6 \\ 
  $\beta_{V}$ & Pine volume, count & 0.17 & 0.014 & 0.14 & 0.20 \\ 
  $\beta_{K}$ & Trees killed, count & -0.0020 & 0.0020 & -0.0071 & -0.000058 \\ 
  $\beta_{BV}$ & Beetle pressure $\times$ pine volume, count & 0.0024 & 0.014 & -0.024 & 0.029 \\ 
  $\beta_{BK}$ & Beetle pressure $\times$ trees killed, count & -0.0022 & 0.0021 & -0.0079 & -0.000069 \\ 
  $\beta_{VK}$ & Volume $\times$ trees killed, count & -0.050 & 0.018 & -0.084 & -0.014 \\ 
  $\tau_{\beta}$ & Year effect scale, count & 0.51 & 0.16 & 0.28 & 0.85 \\ 
  $\sigma_{\beta}$ & GP scale, count & 1.00 & 0.029 & 0.94 & 1.1 \\ 
  $L$ & GP length scale & 11 & 0.34 & 10 & 11 \\ 
  $k$ & Dispersion parameter, count & 3.2 & 0.096 & 3.0 & 3.4 \\ 
   \hline
\end{tabular}
\caption{Parameter estimates from the phenomenological model with interaction effects. The coefficients here apply to the \textit{standardized} versions of the predictors ($B$, $V$, and $K$) so that the coefficient magnitudes can be interpreted as \textit{predictor importance}. The metrics $\text{CI}_{2.5\%}$ \& $\text{CI}_{97.5\%}$ are the bounds of the 95\% credible intervals.} 
\label{tab:pars_int}
\end{table}

\subsection{Model comparisons} \label{Model comparisons}

Likelihood-based methods of model comparison are computationally difficult for models with Gaussian processes (GPs). Model comparison in this context requires computing the marginal likelihood, which involves integrating over all possible functions that could be drawn from the GP prior. Given these limitations, we explored two alternative approaches to evaluate model performance.

The first approach involved comparing non-GP versions of the phenomenological models (i.e, with and without interaction effects) to assess relative performance. The main model significantly outperformed the interaction effect version, despite an expected bias towards complexity. Non-GP models effectively have an inflated sample size because they fail to account for spatial autocorrelation, and high sample sizes favor complex models. This finding suggests that the main model is sufficiently complex for our needs. For code, see {\fontfamily{qcr}\selectfont scripts/model\_diagnostics.Rmd} in the supplementary files.

\begin{figure}[H]
\centering
\includegraphics[scale = 1]{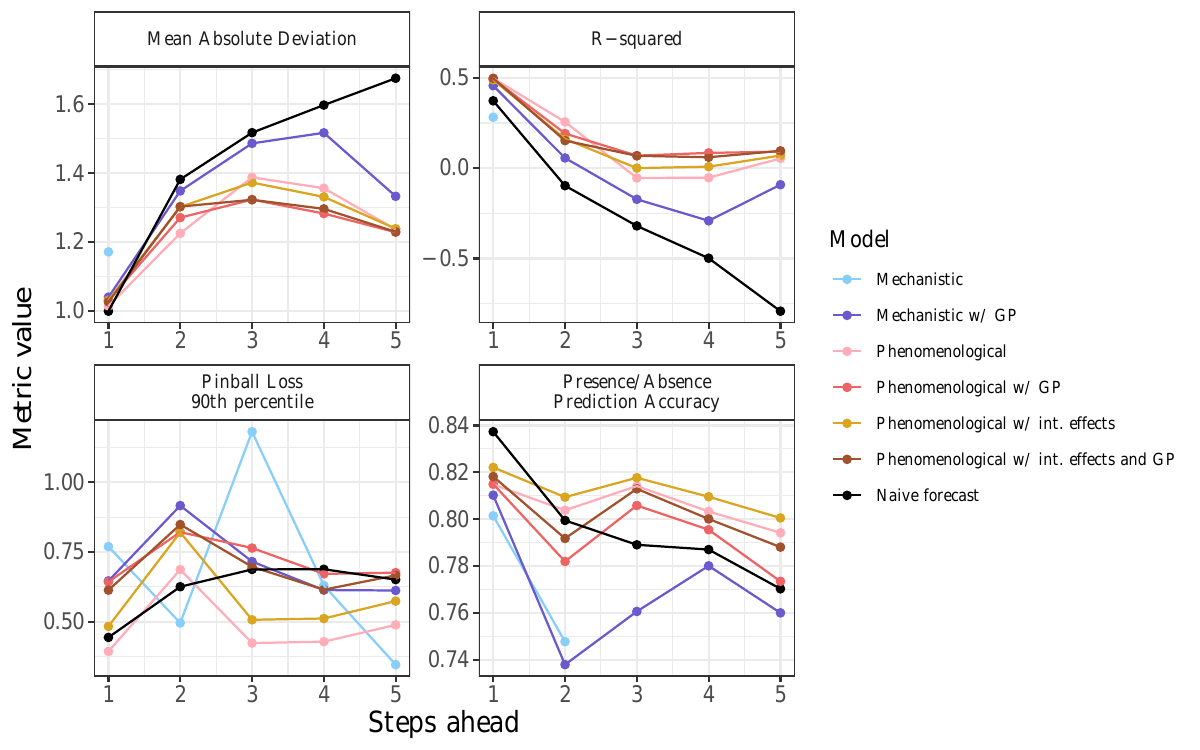}
\caption{Comparison of predictive performance across all models for forecasts up to five years into the future. Results include a na\"{i}ve baseline forecast that assumes next year's value will match the current year. Note that some subplots use a reduced y-axis range to better highlight differences between top-performing models; this only censors some data from the mechanistic model without Gaussian processes. The mean absolute deviation applies to predictions of $\log(I_t(x))$, only for locations with $I_t(x) > 0$. The same applies to R squared and pinball loss. Only the presence/absence prediction accuracy was computed using all cells.}
\label{fig:model_comparison1}
\end{figure}

The second approach utilized simulations to evaluate predictive performance across multiple time horizons. We simulated data using known year effects and overwintering survival values (but without fitted GP realizations). This approach tested in-sample predictive ability. The results are shown in Figure \ref{fig:model_comparison1}.


Phenomenological models generally outperformed mechanistic models across most metrics. However, one-year-ahead predictions showed similar accuracy between both classes of models (e.g., 0.81 vs. 0.82 presence/absence accuracy). The inclusion of GPs proved particularly important for capturing prediction uncertainty. While GP and non-GP versions of the phenomenological models showed similar performance across most metrics, a notable difference emerged in their ability to characterize the range of possible values, as measured by pinball loss. Indeed, the GP models generated wider predictive intervals that nearly encapsulated the actual outbreak trajectory over 11 years (Fig. \ref{fig:time_series_ribbons_combined}). This superior representation of uncertainty highlights the value of incorporating GPs, despite the computational challenges they present for model-fitting and model comparisons.

\subsection{Alternative model results} \label{Alternative model results}

\begin{figure}[H]
\centering
\includegraphics[scale = 1]{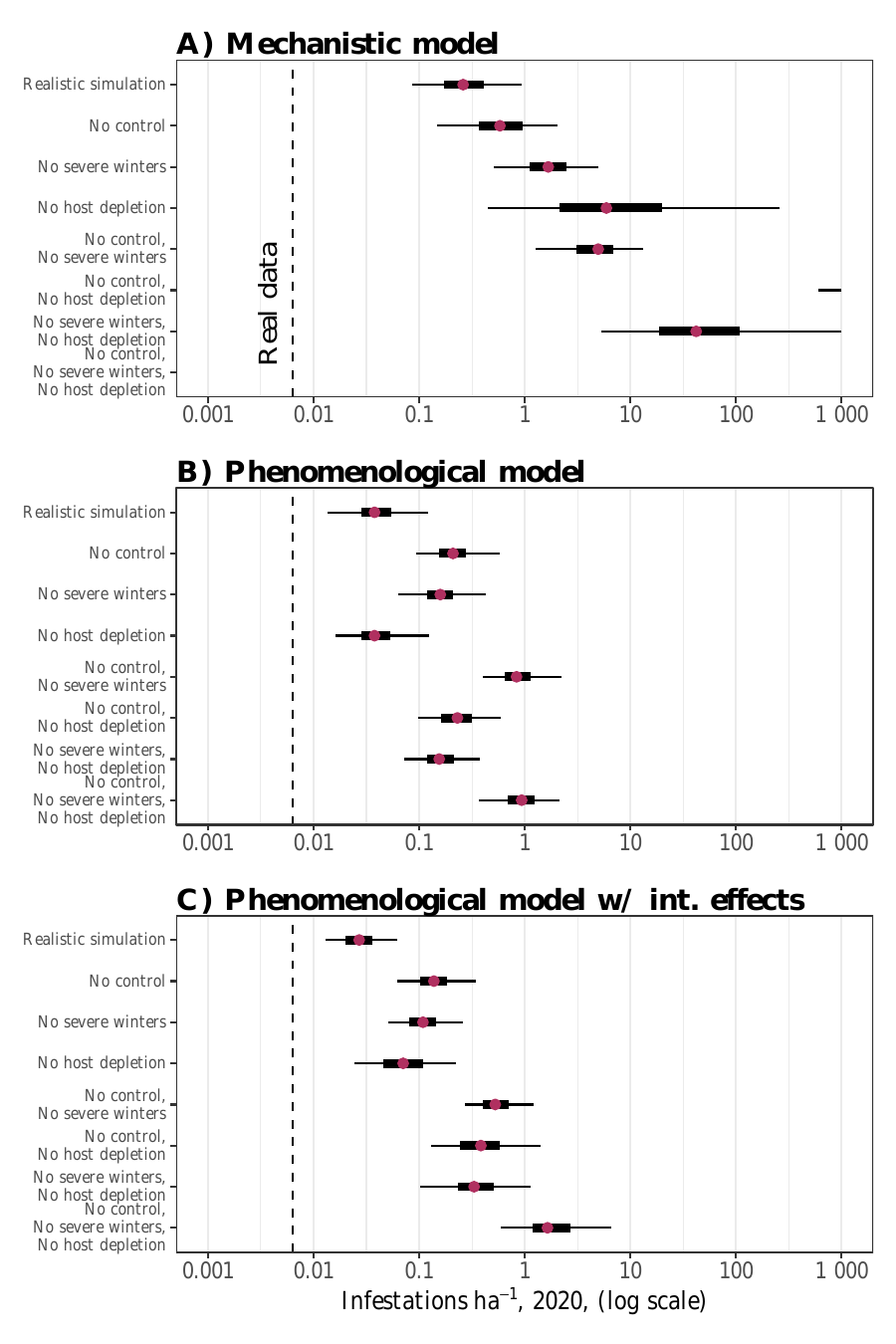}
\caption{Distributions of infestation density in 2020, stratified by simulation scenario and model structure. Points, horizontal thick lines, and thin lines respectively show the mean, 50\% interval, and 95\% interval of the predictive posterior distributions. All values are spatial averages across the study area. Dashed vertical lines show the actual value. Simulations use the observed environmental parameters from 2009--2019, following the \textit{actual outbreak} approach (see Section \ref{Simulations} for details).}
\label{fig:combined_Infest_boxplots}
\end{figure}

\begin{figure}[H]
\centering
\includegraphics[scale = 1]{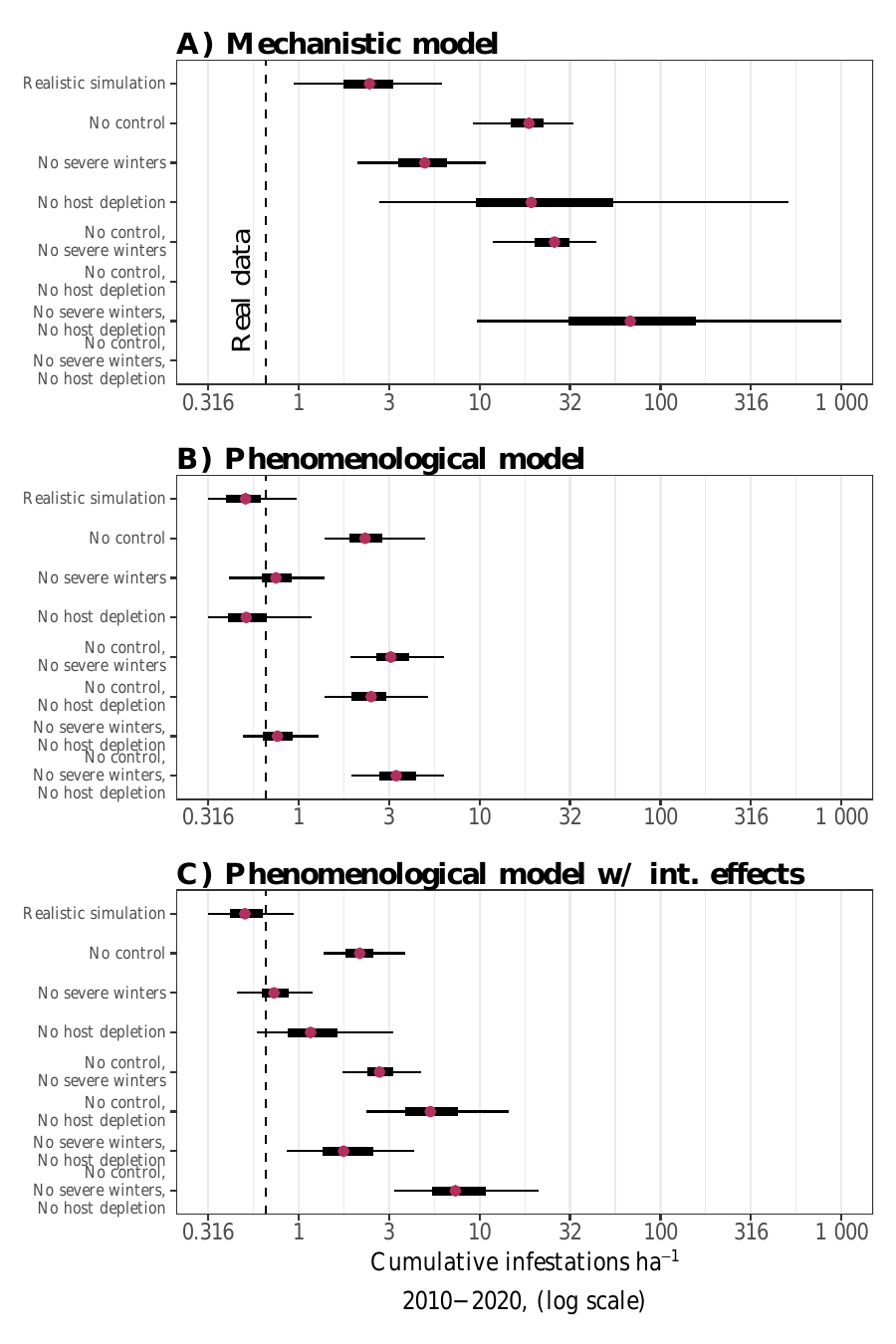}
\caption{Distributions of cumulative infestations from 2010--2020, stratified by simulation scenario and model structure. Points, horizontal thick lines, and thin lines respectively show the mean, 50\% interval, and 95\% interval of the predictive posterior distributions. All values are spatial averages across the study area. Dashed vertical lines show the actual value. Simulations use the observed environmental parameters from 2009--2019, following the \textit{actual outbreak} approach (see Section \ref{Simulations} for details).}
\label{fig:combined_ctk_boxplots}
\end{figure}

\begin{figure}[H]
\centering
\includegraphics[scale = 1]{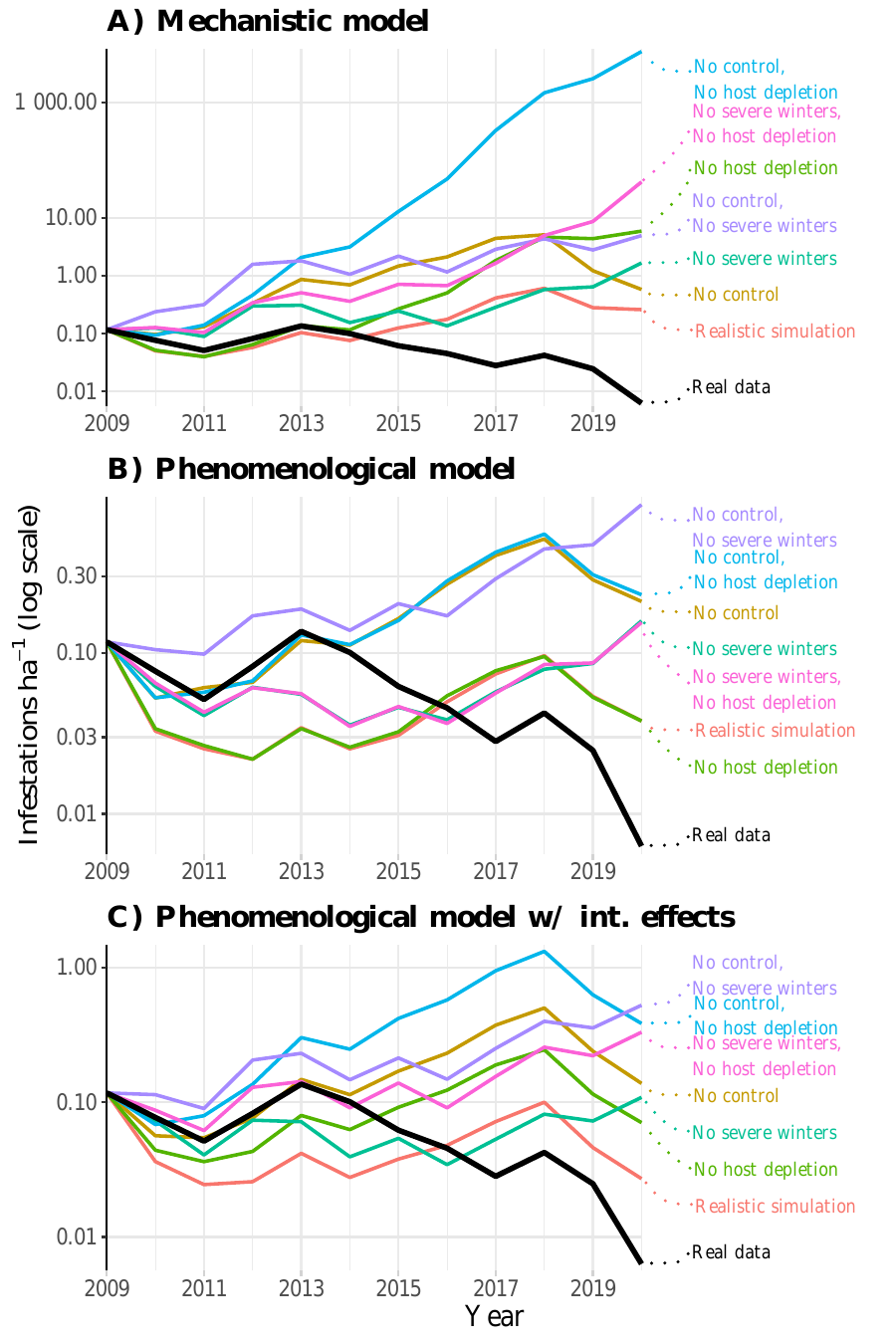}
\caption{Time series of mountain pine beetle infestation densities, in both counterfactual simulation scenarios (colored lines show the median value across simulations), and in the real data (black line). Simulations use the observed environmental parameters from 2009--2019, following the \textit{actual outbreak} approach (see Section \ref{Simulations} for details). The y-axis values represent the spatial average of infestations $\text{ha}^{-1}$ across the study area.}
\label{fig:combined_TS}
\end{figure}

The alternative models confirmed two main findings of the main model: control limited total tree damage, and the outbreak collapsed due to a combination of control efforts and cold winters. However, there were some differences. The mechanistic model predicts higher average infestation densities and greater variability across simulations, and both alternative models suggest a small effect of host-tree depletion. To visualize similarities and differences between the three models, we re-created Figures and Tables from the main text, but with separate panels or rows for each model (Figures \ref{fig:combined_ctk_boxplots}, \ref{fig:combined_Infest_boxplots}, \ref{fig:combined_TS}; Table \ref{tab:control_delta2}).

The mechanistic model generally predicts higher infestation densities. This is especially true in no-control simulation scenarios. The mechanistic model generates a wider range of outcomes, both in terms of the final (2020) infestation density (Fig.\ref{fig:combined_Infest_boxplots}, panel A), and the cumulative number of killed trees (Fig. \ref{fig:combined_ctk_boxplots}, panel A). For example, in the no host-tree depletion scenario, the cumulative number of killed trees ranges from 3 to 500 trees per hectare (Fig. \ref{fig:combined_ctk_boxplots}, panel A, row 4).

The mechanistic model infers a higher relative effectiveness of control, defined as the proportion of trees saved over an 11-year period compared to a do-nothing scenario. Specifically, the mechanistic model predicts that 87\% of trees are saved (95\% PI: 71--94\%), whereas the main model predicts 79\% (95\% PI: 58--89\%).

We suspect that the different quantitative predictions of the mechanistic model can be attributed to how population growth dynamics are represented. In the mechanistic model, population growth is explicitly tied to beetle pressures through the relationship $\mathrm{E}\left[\log(I_t(x)) | I_t(x) > 0\right] = c_t \, b_t(x) \, s_t(x)$, whereas the phenomenological model incorporates intercept parameters $\gamma_0$ and $\beta_0$ that moderate this dependence. It is unclear what model structure is better \textit{a priori}. The mechanistic model explicitly encodes geometrics growth, and while beetles exhibit geometric growth at local scales, the effective dynamics might look very different at the scale of a 5$\times$5 km cell.

Host-tree depletion played a significant role in the mechanistic model (Fig. \ref{fig:combined_TS}, panel A), but this was due to specific modeling assumptions rather than a data-driven inference. This assumption in question is a moderately-informative prior on the $\alpha$ parameter, which was necessary for estimation since $\lambda_1$ and $\alpha$ are otherwise effectively non-identifiable. The posterior contraction for $\alpha$ indicates substantial prior influence on the posterior distribution (see Appendix \ref{model_details}). When the $\alpha$ is small, the initial number of susceptible trees is low, which eventually constrains infestation densities. The phenomenological model with interaction effects exhibits an effect of host-tree depletion (Fig. \ref{fig:combined_TS}, panel C), and this is not due to modeling assumptions. On a technical level, this can be attributed to the negative interaction effect coefficient $\gamma_{BK}$ (Table \ref{tab:pars_int}).

\begin{figure}[H]
\centering
\includegraphics[scale = 1]{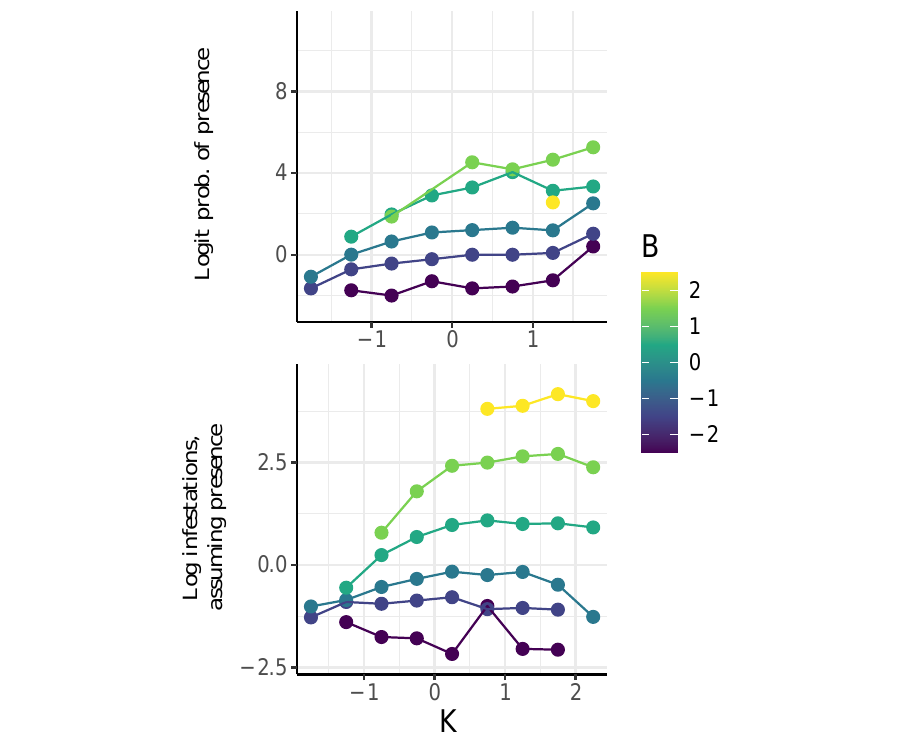}
\caption{There is no clear relationship between infestations and the cumulative killed trees. Points represent average estimates within evenly spaced bins across the standardized $K$ and $B$ predictors. We stratify by the beetle pressure predictor to avoid a confounding bias, since areas with higher tree mortality tend to have more beetles and thus higher infestation rates.}
\label{fig:htd}
\end{figure}

Despite the interaction effect model showing a small effect of host-tree depletion, we believe that the best overall inference is that there is a negligible effect of host-tree depletion via MPB infestations (recall that this paper does not consider depletion via logging activity). The first reason is that the main phenomenological model is superior according to cross-validation, and this model implies a negligible effect. The second reason is that host-tree depletion is rather low in absolute terms: average infestation densities in Alberta are 2-3 orders of magnitude less than in British Columbia. The third reason is graphical evidence: there is no clear relationship between the cumulative killed trees predictor $K$ and MPB infestations, even when we stratify by the beetle pressure predictor $B$ (Fig. \ref{fig:htd}).

\renewcommand*\rot{\multicolumn{1}{R{30}{1em}}}
\renewcommand\cellset{\renewcommand\arraystretch{0.8}%
\setlength\extrarowheight{0pt}}
\begin{table}[H]
\begingroup\fontsize{9pt}{10pt}\selectfont
\begin{tabular}{lllllll}
\rot{Simulation scenario} & \rot{Model} & \rot{Trees Killed $\text{ha}^{-1}$ (No control)} & \rot{Trees killed $\text{ha}^{-1}$ (Control)} & \rot{No. trees $\text{ha}^{-1}$ Saved by control} & \rot{\% Trees Saved by control} & \rot{Control multiplier} \\ 
  \hline \\
\makecell[tr]{Actual outbreak,\\2010--2020} & \makecell[tr]{Mechanistic} & 19 (9.2-33) & 2.5 (0.94-6.2) & 16 (7-29) & 87 (71-94) & 15 (5.7-38) \\ 
  \makecell[tr]{Actual outbreak,\\2010--2020} & \makecell[tr]{Phenomenological} & 2.3 (1.4-5) & 0.51 (0.26-0.98) & 1.8 (0.91-4.1) & 79 (58-89) & 6.7 (2.6-15) \\ 
  \makecell[tr]{Actual outbreak,\\2010--2020} & \makecell[tr]{Phenomenological\\w/ interaction effects} & 2.2 (1.4-3.9) & 0.51 (0.3-0.94) & 1.6 (0.76-3.2) & 76 (57-88) & 5.9 (2.6-12) \\ 
  \makecell[tr]{Future outbreak,\\11 year period} & \makecell[tr]{Mechanistic} & 13 (1.1-40) & 2.5 (0.14-13) & 10 (0.9-30) & 81 (55-96) & 9 (2.5-40) \\ 
  \makecell[tr]{Future outbreak,\\11 year period} & \makecell[tr]{Phenomenological} & 2.2 (0.41-9.7) & 0.55 (0.089-2.4) & 1.5 (0.26-8) & 74 (43-88) & 4.9 (1.1-13) \\ 
  \makecell[tr]{Future outbreak,\\11 year period} & \makecell[tr]{Phenomenological\\w/ interaction effects} & 2 (0.41-8.8) & 0.58 (0.11-2.2) & 1.4 (0.21-6.7) & 71 (42-86) & 4.4 (1.3-11) \\ 
   \hline
\end{tabular}
\endgroup
\caption{How many trees did control efforts save over an 11-year period? \textit{No control} vs. \textit{Control} columns correspond to a control efficacy of $m= 0$ vs. $m=0.47$ control; units are trees per hectare, averaged across the study area. The format of table entries is \textit{Median (95\% predictive intervals)}. The \textit{actual outbreak} simulation scenario uses estimated values of overwintering survival and year effects. The \textit{Future outbreak} scenario uses bootstrapped maps of overwintering survival and randomly generated year effects. The \textit{control multiplier} is calculated as the number of trees saved by control from 2010--2020, divided by the number of controlled trees from 2009--2019.} 
\label{tab:control_delta2}
\end{table}

\section{Additional figures and tables} \label{Supplementary figures and tables}

\begin{figure}[H]
\centering
\includegraphics[scale = 1]{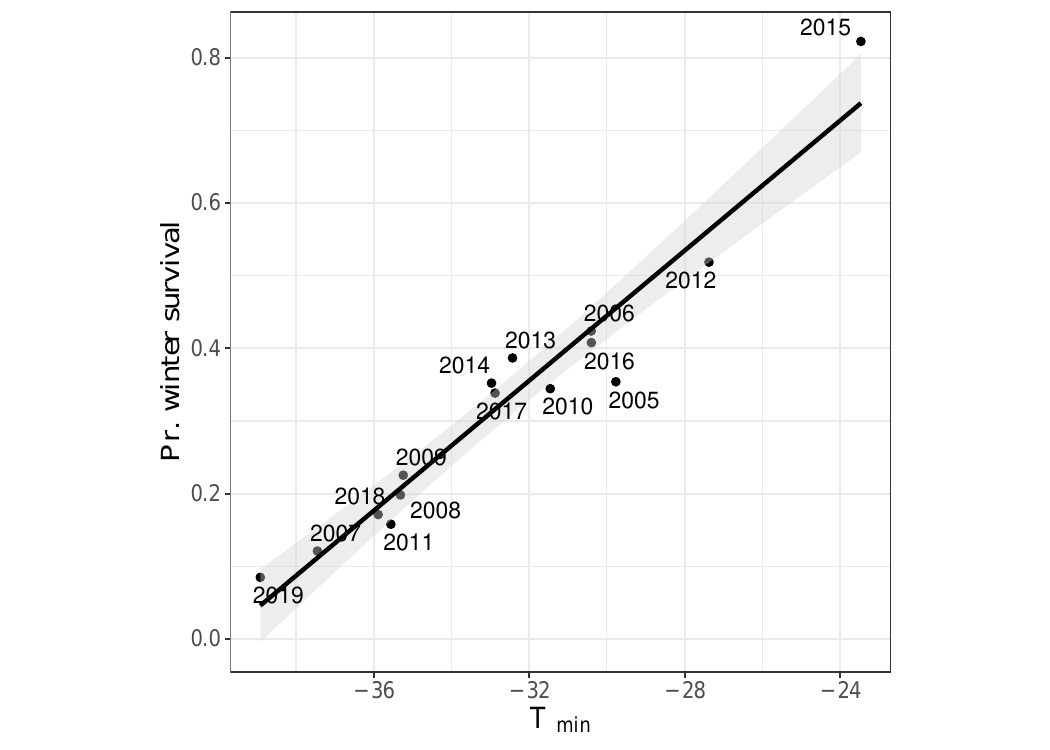}
\caption{The probability of overwintering survival is an approximately linear function of minimum winter temperature ($\text{T}_{\text{min}}$; units: degrees Celcius). Each point shows the average value across all cells that had at least one infestation during the study period.} 
\label{fig:Psurv_vs_Tmin}
\end{figure}

\begin{figure}[H]
\centering
\includegraphics[scale = 1]{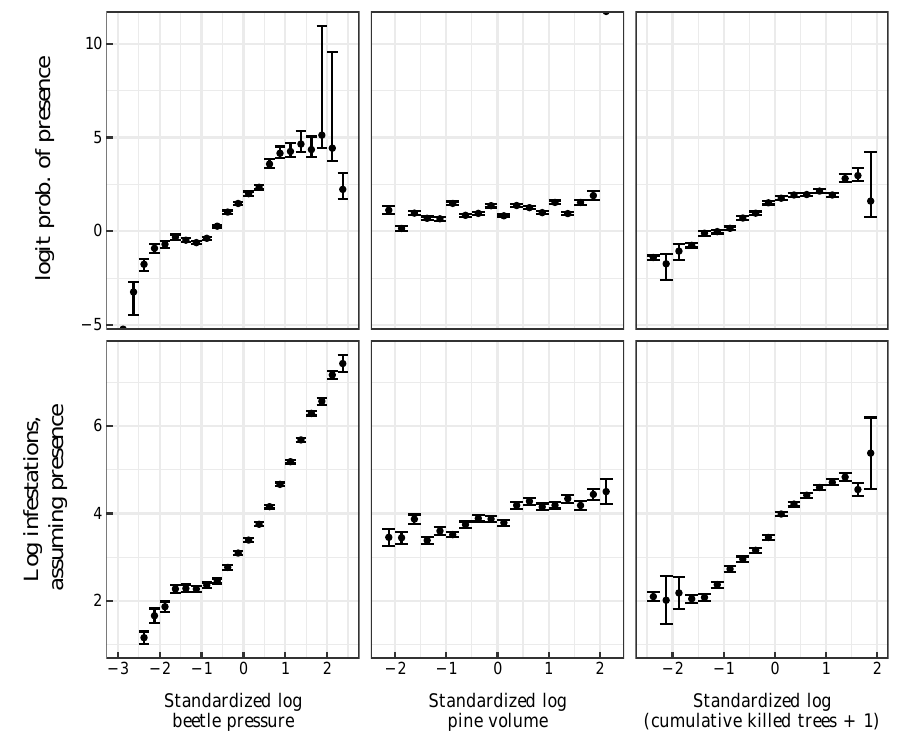}
\caption{Evidence for linear effects of the predictors $B$, $V$, and $K$. Points and bar represent means $\pm$ 1 standard error within evenly-spaced intervals of predictor values.}
\label{fig:mod2_just}
\end{figure}

\begin{figure}[H]
\centering
\includegraphics[scale = 1]{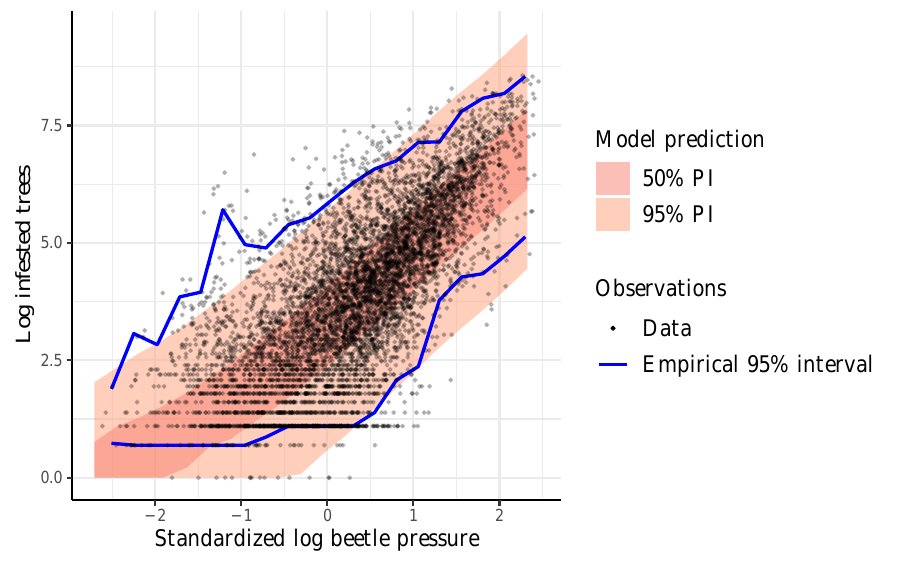}
\caption{The negative binomial distribution captures the observed variability in infestation densities. Each point is a particular cell in a particular year, excluding cells with no infestations. The empirical 95\% interval was calculated with the empirical cumulative distribution function of the cells within evenly-spaced intervals of x-axis values.}
\label{fig:NB_pred_vs_obs}
\end{figure}

\begin{figure}[H]
\centering
\includegraphics[scale = 1]{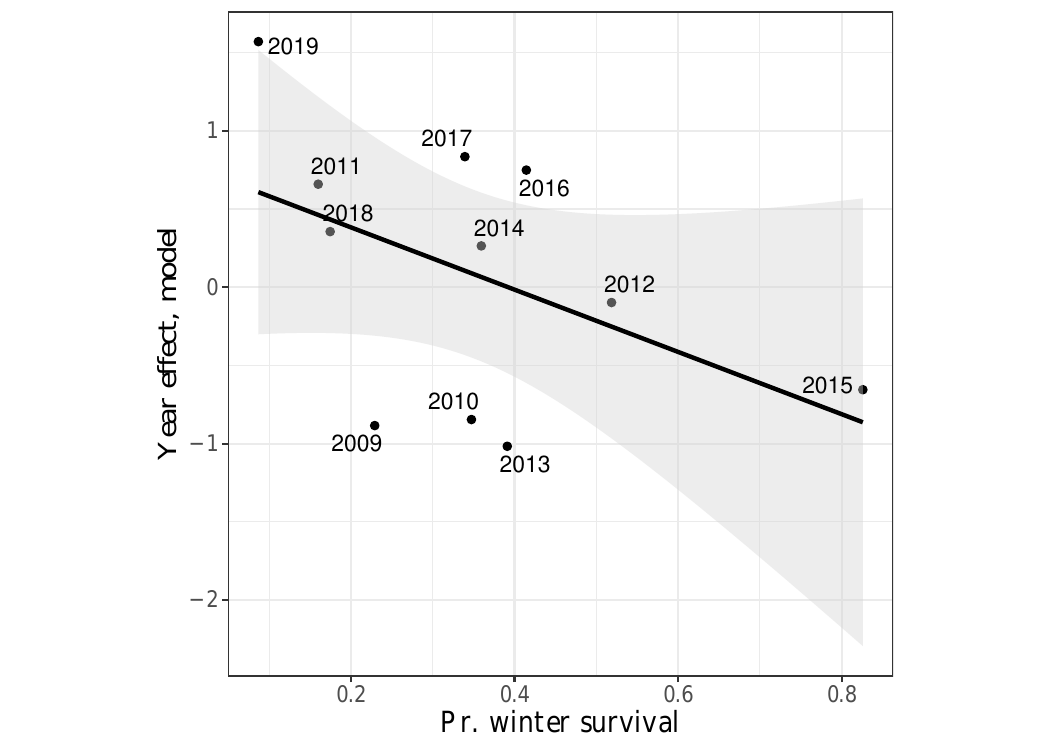}
\caption{There is not a clear relationship between year effect estimates (posterior means) and the probability of winter survival (average across cells with at least one infestation throughout the study period). }
\label{fig:Psurv_vs_c}
\end{figure}

\begin{figure}[H]
\centering
\includegraphics[scale = 1]{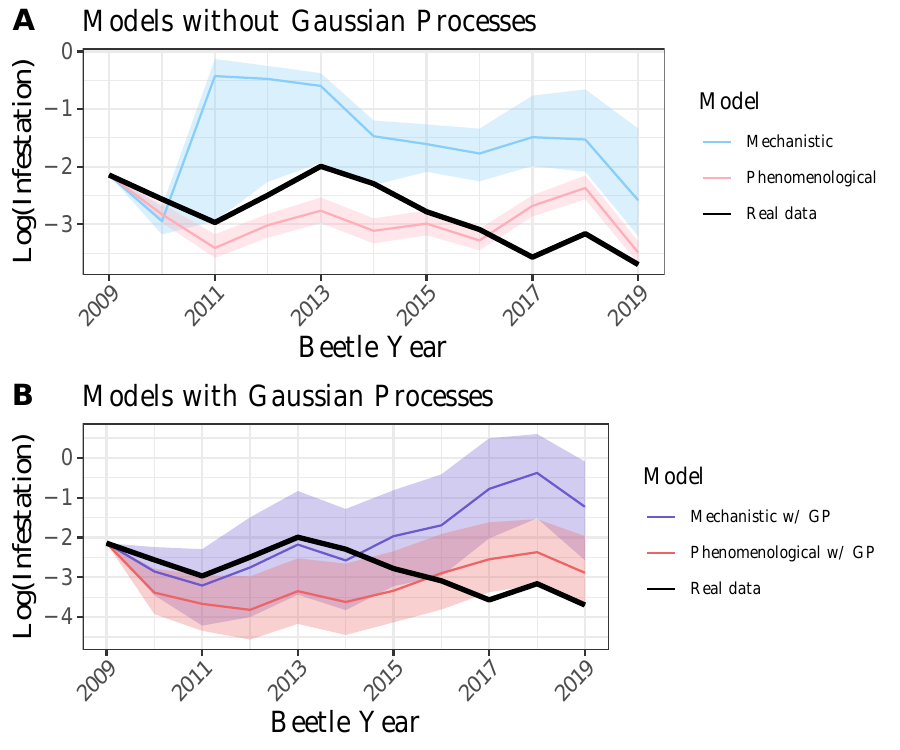}
\caption{Models with Gaussian processes predict a wider range of outbreak trajectories. Colored ribbons show the 95\% predictive intervals of log infestations per hectare, averaged across the study area. Simulations use the \textit{actual outbreak} scenario.}
\label{fig:time_series_ribbons_combined}
\end{figure}

\begin{figure}[H]
\centering
\includegraphics[scale = 1]{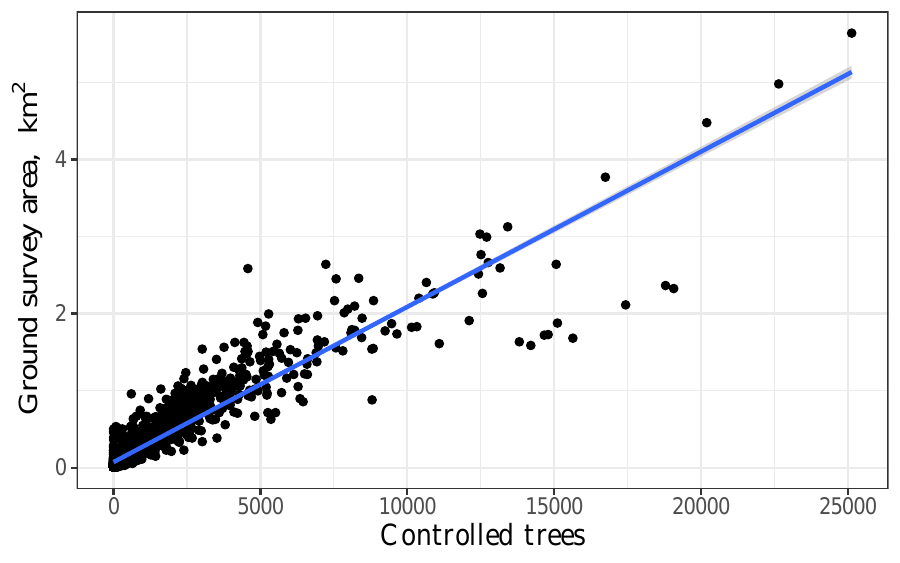}
\caption{The number of trees controlled with level 1 treatment increases linearly with the total area covered by ground surveys. Each point represents the sum of controlled trees within a 10$\times$10 km cell, in a particular year.}
\label{fig:area_vs_control}
\end{figure}






\section{Model-fitting details} \label{model_details}

We examined a suite of diagnostics for Hamiltonian Monte Carlo \citep[Ch. 6]{gelman2014bayesian}. All models under consideration passed standard diagnostic thresholds. Specifically, we confirmed that $\hat{R} < 1.1$ for all parameters (indicating proper chain mixing), the effective sample size per iteration exceeded 0.001 (demonstrating efficient sampling), the energy Bayesian fraction of missing information (E-BFMI) was below 0.2 (suggesting appropriate model specification), and the proportion of divergent trajectories remained well below 1\% (indicating unbiased estimation). The diagnostic analysis can be found in the supplementary files, specifically {\fontfamily{qcr}\selectfont scripts/model\_diagnostics.Rmd \& \selectfont scripts/stan\_utility.R}

To evaluate the influence of the prior distributions, we calculated the posterior contraction:
\begin{equation}
\text{post. contraction} = 1 - \frac{ \mathbb{V}{\mathrm{post}}} { \mathbb{V}{\mathrm{prior}}}.
\end{equation}
This metric quantifies how informative the data are relative to the prior for each parameter. All parameters showed posterior contraction greater than 0.99, with the exception of parameter $\alpha$ in the mechanistic model, whose posterior contraction was approximately 0.5. The posterior contraction calculations can also be found in the supplementary files: {\fontfamily{qcr}\selectfont  scripts/model\_diagnostics.Rmd}.

\end{appendices}

\bibliographystyle{apalike}
\bibliography{mpb_refs.bib}

\end{document}